\documentclass[12pt,onecolumn,draft]{IEEEtran}
\usepackage{times}
\usepackage[final]{graphicx}
\usepackage{amsmath,amsfonts}
\usepackage{amssymb,amsbsy}
\usepackage{times}

\usepackage{epsf}
\usepackage{cite}

\newtheorem{lem}{Lemma}
\newtheorem{rem}{\textbf{Remark}}

\newtheorem{prp}{\textbf{Proposition}}
\newtheorem{thm}{\textbf{Theorem}}
\newtheorem{cor}{\textbf{Corollary}}

\newcommand{\ep}{\epsilon}

\newcommand\ignore[1]{}

\newcommand{\hsp}{\hspace{0.05in} }

\newcommand{\hspp}{\hspace{0.02in} }
\newcommand{\hsppp}{\hspace{0.01in} }
\newcommand{\ee}{ {\sf E} }

\newcommand{\bH}{\mathbf{H}  }
\newcommand{\bQ}{\mathbf{Q}}
\newcommand{\bEe}{{\mathbf{E}}}

\newcommand{\by}  { {\mathbf{y}}  }

\newcommand{\bx}  { {\mathbf{x}}  }

\newcommand{\ud} { {\mathrm{d}}  }  
\newcommand{\bw} { {\mathbf{w}} }
\newcommand{\bI} { {\mathbf{I}} }

\newcommand{\snr}{{\sf{SNR}} }

\newcommand{\ord}{{\mathcal{O}}}

\newcommand{\diag} {{\mathrm{diag}}  }

\newcommand{\Tcoh}{\mathnormal{T_{coh}}}
\newcommand{\Wcoh}{\mathnormal{W_{coh}}}
\newcommand{\Ncoh}{\mathnormal{N_{c}}}

\newcommand{\nc}{\mathnormal{N_{c}}}
\newcommand{\po}{\mathnormal{P_{0}}}
\newcommand{\poc}{\mathnormal{P_{0,\sf c}}}
\newcommand{\ponc}{\mathnormal{P_{0, \sf nc}}}
\newcommand{\De}{ {\sf D}_{ \sf eff } } 

\newcommand{\peaky}{{\mathrm{peaky}} }

\newcommand{\fone}{f_1}
\newcommand{\ftwo}{f_2}
\newcommand{\fthr}{f_3}
\newcommand{\gone}{g_1}
\newcommand{\gtwo}{g_2}

\newcommand{\Dwmax} { D_{W,\hspp \max} }
\newcommand{\Dtmax} { D_{T,\hspp \max} }

\newcommand{\etas} {\eta^{*}}

\newcommand{\prob}{ {\mathrm{Pr}} }
\newcommand{\HT}{ {\sf h_t} }
\newcommand{\HTtr}{ {\sf h^{train}_{t}} }
\newcommand{\HTtrs}{ {\sf h^{train, \hspp \star}_{t}} }

\newcommand{\coh}{ {\sf coh} }
\newcommand{\bX} { {\bf X} }

\newcommand{\inst} { {\sf inst} }
\newcommand{\train} { {\sf train} }

\newcommand{\LT}{{\sf LT} }
\newcommand{\ST}{{\sf ST} }
\newcommand{\ca}{{\sf c}}
\newcommand{\nca}{{\sf nc}}

\newcommand{\tsty} {\textstyle}

\begin{document}
\renewcommand{\textfraction}{0}

\title{ 
Capacity of Sparse Wideband Channels with 
\\ [1.5 mm]
Partial Channel Feedback
\large \author{Gautham Hariharan, Vasanthan Raghavan and Akbar M.\ Sayeed 
\thanks{This work was supported in part by the NSF through grant \#CCF-0431088. 
G.\ Hariharan is with the Qualcomm Inc., San Diego, CA 92121, USA 
({\tt{gharihar@qualcomm.com}}). V.\ Raghavan is with the Coordinated Science 
Laboratory and the Department of Electrical and Computer Engineering, University 
of Illinois at Urbana-Champaign, Urbana, IL 61801, USA 
({\tt{vasanthan\_raghavan@ieee.org}}). 
A.\ M.\ Sayeed is with the Department of Electrical and Computer Engineering, 
University of Wisconsin-Madison, Madison, WI 53706, USA ({\tt{akbar@engr.wisc.edu}}). 
  }}}

\date{}
\maketitle

\begin{abstract}
\noindent 
This paper studies the ergodic capacity of wideband multipath channels 
with limited feedback. Our work builds on recent results that have established 
the possibility of significant capacity gains in the wideband$/$low-$\snr$ 
regime when there is perfect channel state information (CSI) at the transmitter.
Furthermore, the perfect CSI benchmark gain can be obtained with the feedback 
of just one bit per channel coefficient. However, the input signals used in these 
methods are peaky, that is, they have a large peak-to-average power ratios. 
Signal peakiness is related to channel coherence and many recent measurement 
campaigns show that, in contrast to previous assumptions, wideband channels 
exhibit a sparse multipath structure that naturally leads to coherence in time 
and frequency. In this work, we first show that {\em even} an instantaneous power 
constraint is sufficient to achieve the benchmark gain when perfect CSI is 
available at the receiver. In the more realistic non-coherent setting, we study 
the performance of a training-based signaling 
scheme. We show that multipath sparsity can be 
leveraged to achieve the benchmark gain under both average as well as instantaneous 
power constraints as long as the channel coherence scales at a sufficiently 
fast rate with signal space dimensions. We also present rules of thumb on choosing 
signaling parameters as a function of the channel parameters so that the full benefits 
of sparsity can be realized. 
\end{abstract}

\section{Introduction}
\label{sec1} 
Recent research on the fundamental limits of wideband$/$low-$\snr$ communications 
has focused on the non-coherent regime where the impact of channel state information 
(CSI) on the achievable rates is critical. From a capacity perspective, spreading 
signals has been shown to be sub-optimal~\cite{medard_gallager} and peaky or flash 
signaling schemes are necessary~\cite{verdu,telatar_tse} to achieve the non-coherent 
wideband capacity. Recent work by 
Zheng {\emph{et al.}}~\cite{zheng} has emphasized the crucial role of channel 
coherence in the low-$\snr$ regime and the importance of implicit$/$explicit 
channel learning
schemes that can bridge the gap between the coherent and the 
non-coherent extremes. However, these results have been derived based 
on an implicit assumption of rich multipath where the number of independent 
degrees of freedom (DoF) in the delay domain scale linearly with bandwidth. 

Recent measurement campaigns in the case of ultrawideband systems show that 
the number of independent DoF do not scale linearly with 
bandwidth~\cite{molisch,saadane,chiachin_residential,porrat2,karedal,molisch_etal,chiachin1}. 
In fact, the physical layer channel model proposed by the IEEE 802.15 working 
group for ultrawideband communication systems exhibits sparsity in the delay 
domain 
(see for example, the measurement data in~\cite[p.\ 15]{ieee802_15_foerster}). 
Motivated by these works, we introduced the notion of \emph{multipath sparsity} 
in~\cite{capacity_sparse_jstsp} as a source of channel coherence and proposed a 
channel modeling framework to capture the impact of sparsity in delay and Doppler 
on achievable rates. 
The analysis in~\cite{capacity_sparse_jstsp} shows that multipath 
sparsity can help in reducing$/$eliminating the need for peaky signaling in achieving
wideband capacity.

In this work, we build on the results in~\cite{capacity_sparse_jstsp} and study 
the impact of {\em channel state feedback} on achievable rates in 
sparse wideband channels. Although earlier works (for
example~\cite{goldsmith_varaiya,caire,proakis_tut} and references
therein) have explored capacity with transmitter CSI, it is only
recently~\cite{verdu,borade,manish} that the impact of feedback in the
low-$\snr$, non-coherent regime has received attention. In
particular, in the low-$\snr$ regime, it is shown in~\cite{borade,verdu} that 
with an average power constraint, the capacity gain with perfect transmitter 
and receiver CSI 
over the case when there is only perfect receiver CSI 
is $\log \left(\frac{1}{\snr} \right)$. 
More interestingly, it is shown that a {\em limited feedback} scheme where 
only one bit per independent DoF is available at the transmitter can also 
achieve a gain of $\log \left( \frac{1}{\snr} \right)$~\cite{borade,verdu}. 
However, for both the optimal waterfilling scheme~\cite{gallager_book,goldsmith_varaiya} 
as well as the one bit limited feedback scheme, the input signal tends to be 
peaky (or bursty) in time, leading to a high peak-to-average power ratio, and
difficulties from an implementation standpoint. The need to reliably estimate 
the channel at the receiver leads to the use of peaky training followed by 
communication in~\cite{borade}. Similar results have also been reported 
in~\cite{manish} where the authors study the optimization of the training 
length, average training power and spreading bandwidth in a wideband setting. 

The focus of this work is on leveraging multipath sparsity to overcome or reduce 
the need for peaky signaling schemes. We work towards this goal by providing 
a concise description of the sparse channel model~\cite{capacity_sparse_jstsp} 
in Sec.~\ref{sec2}. We then study the performance in the 
case where the receiver has perfect 
CSI and the transmitter has one bit (per independent DoF) in Sec.~\ref{sec3}. 
In contrast to~\cite{verdu,borade,manish} which study the performance only 
under an {\em average} (or long-term) power constraint, we also consider an 
{\em instantaneous} (or short-term) power constraint. We restrict our attention 
to {\em causal} signaling schemes that can be realized in practice. We show that 
an optimal threshold of the form $\HT = \lambda \log \left(\frac{1}{\snr} \right)$ 
for any $\lambda \in (0,1)$ provides a measure of achievable 
rate\footnote{All logarithms are assumed to be base $e$ and the units for all 
rate quantities are assumed to be nats per channel use.} which 
behaves as $\left(1+\HT \right) \snr$ in the wideband limit. Thus when 
$\lambda$ approaches $1$, we achieve the perfect transmitter 
CSI capacity which is the benchmark for all limited feedback schemes. We derive 
a sufficient condition under which this benchmark can be approached even with an 
instantaneous power constraint. A key parameter that determines this condition is 
$\bEe \left[ \De \right]$, the average number of {\em active} 
independent channel dimensions, the number of independent channel 
coefficients that exceed the threshold in the power allocation 
scheme. In particular, with an instantaneous power constraint, the 
benchmark capacity gain is achieved when $\bEe \left[ \De \right] - \HT 
\rightarrow \infty$ as $\snr \rightarrow 0$. We discuss the 
feasibility of the above condition when the channel is rich as well as sparse. 

In Sec.~\ref{sec4}, the focus is on the case where the receiver has no 
CSI {\em a priori} and a training-based signaling 
scheme is employed. Along the same 
lines as in~\cite{borade,manish}, we study the rates achievable with this scheme, 
albeit for sparse channels. With an average power constraint, it is
shown that as long as the channel coherence dimension $\nc$ scales
with $\snr$ as $\nc = \frac{1}{\snr^{\mu}}$ for some $\mu
> 1$, the rate achievable with the training 
scheme converges to the capacity with perfect transmitter CSI, the performance 
benchmark, in the wideband 
limit. Furthermore, this condition is achievable only when the
channel is sparse and we provide guidelines on choosing the signal
space parameters (signaling$/$packet duration, bandwidth and transmit
power) such that $\mu > 1$ is realized. The critical role of
channel sparsity is further revealed when we impose an
instantaneous power constraint. In contrast to peaky signaling
that violates the finiteness constraint on the peak-to-average power,
channel sparsity is necessary to realize the conditions required to 
approach the performance gain with an instantaneous power constraint: 
$\mu > 1$ and $\bEe \left[ \De \right] - \HT \rightarrow \infty$. We 
summarize the paper in Sec.~\ref{sec5} by highlighting our contributions 
and placing them in the context of~\cite{verdu,borade,manish}.

\section{System Model}
\label{sec2} 
In this section, we elucidate the model
developed in~\cite{capacity_sparse_jstsp} for sparse multipath
channels. Our results are based on an orthogonal short-time
Fourier (STF) signaling framework~\cite{ke_tamer_say,kozek} that
naturally relates multipath sparsity in delay-Doppler to coherence
in time and frequency.

\subsection{Sparse Multipath Channel Modeling}
\label{sec2a}
A discrete, physical multipath channel can be modeled as
\begin{eqnarray}
y(t) & = & \int_{0}^{T_m} \! \! \!
\int_{-\frac{W_d}{2}}^{\frac{W_d}{2}}
 h(\tau,\nu) x(t-\tau) e^{j 2 \pi \nu t} \, \ud \nu \, \ud \tau
 + w (t) \\ 
h(\tau,\nu) & = & \sum_{n} \beta_n \delta(\tau-\tau_n) \delta(\nu - \nu_n) , 
{\hspace{0.2in}} y(t) = \sum_n \beta_n x(t-\tau_n)e^{j2\pi \nu_n t} + w(t)
\label{disc_mp}
\end{eqnarray}
where $h(\tau,\nu)$ is the delay-Doppler spreading function of the
channel, $\beta_n$, $\tau_n \in [0,T_m]$ and $\nu_n \in
[-W_d/2,W_d/2]$ denote the complex path gain, delay and Doppler
shift associated with the $n$-th path. $T_{m}$ and $W_{d}$ denote
the delay and the Doppler spreads, respectively. The quantities $x(t),
y(t)$ and $w(t)$ denote the transmitted, received and additive
white Gaussian noise waveforms, respectively. Throughout this
paper, we assume an {\emph{underspread}} channel where $T_{m}W_{d} \ll
1$.

\begin{figure*}[hbt!]
\begin{center}
\begin{tabular}{ccc}
\begin{minipage}{2.8in}
\centerline{\includegraphics[width=2.8in]{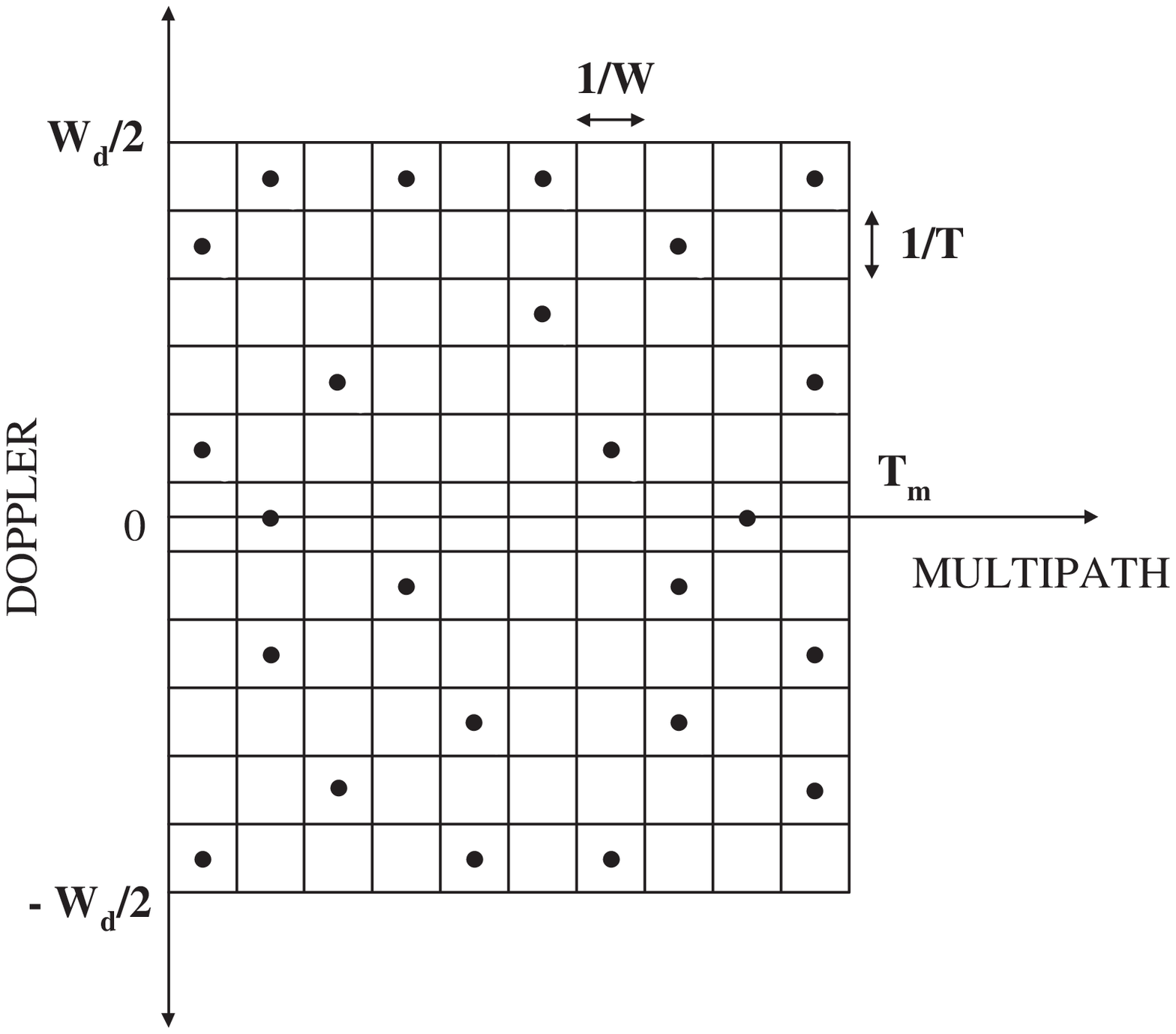}}
\end{minipage} &
\begin{minipage}{3.0in}
\centerline{\includegraphics[width=3.0in]{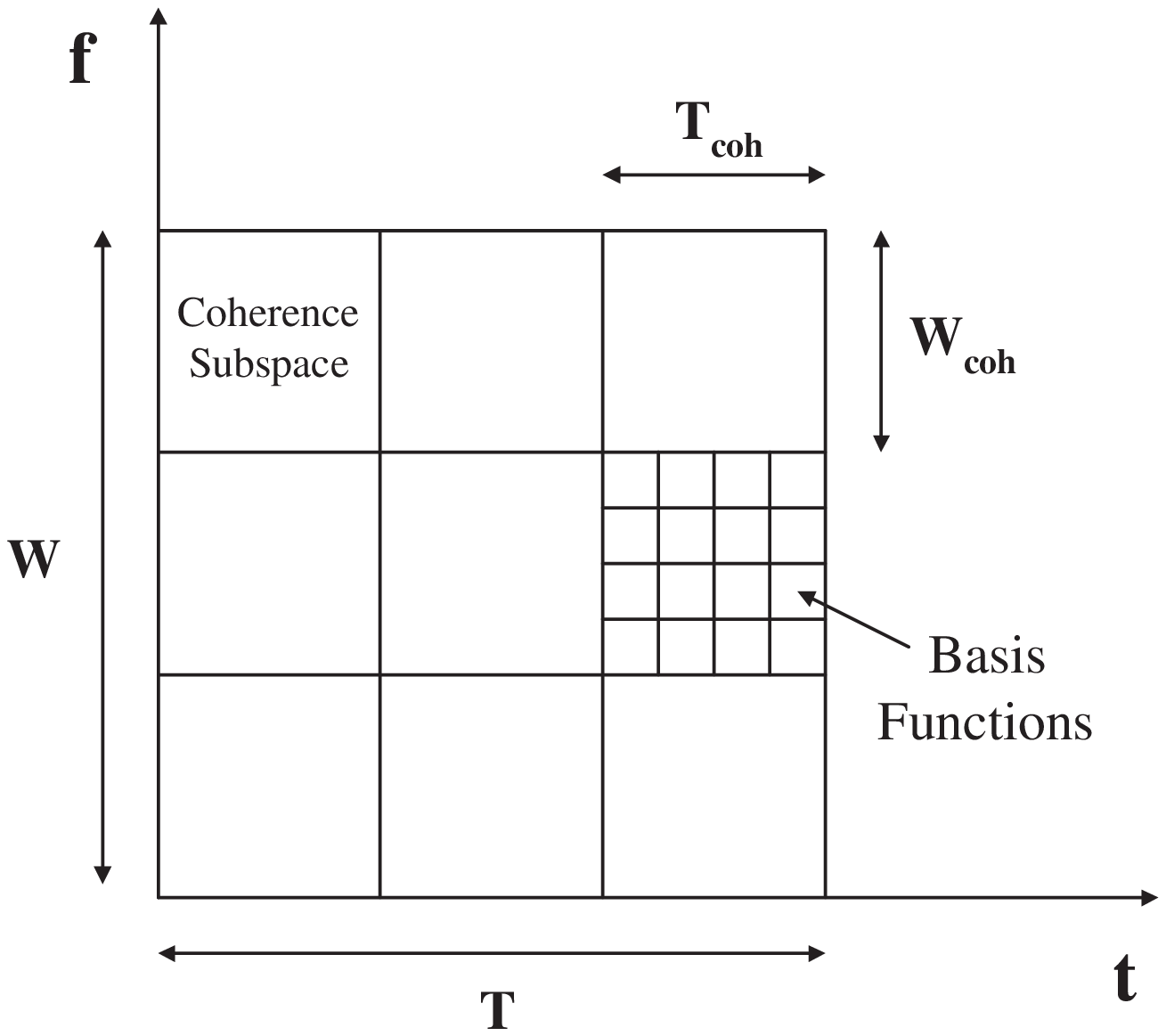}}
\end{minipage} \\
 (a) & (b)
\end{tabular}
\caption{ \label{fig:del_dopp} {\sl (a) Delay-doppler sampling
commensurate with signaling bandwidth and duration. (b)
Time-frequency coherence subspaces in STF signaling.}}
\end{center}
\vspace{-5mm}
\end{figure*}

We use a \emph{virtual
representation}~\cite{akbar_behnaam,akbar_and_venu} of the
physical model in (\ref{disc_mp}) that captures the channel
characteristics in terms of {\em resolvable paths} and greatly
facilitates system analysis from a communication-theoretic
perspective. The virtual representation uniformly samples the
multipath in delay and Doppler at a resolution commensurate with
signaling bandwidth $W$ and signaling duration $T$, respectively. Thus, we have 
\begin{eqnarray}
y(t) &= &\sum_{\ell=0}^{L} \sum_{m=-M}^{M} h_{\ell,m} x(t-
\ell/W)e^{j2\pi
mt/T} + w(t)  \label{del_dopp_samp} \\
 h_{\ell,m} &\approx & \sum_{n \in S_{\tau,\ell} \hspp  \cap \hspp 
S_{\nu,m}}
\beta_n  \label{hlm}
\end{eqnarray}
where $L = \lceil T_m W\rceil$ and $M = \lceil T W_d/2\rceil$. The
sampled representation (\ref{del_dopp_samp}) is linear and is
characterized by the virtual delay-Doppler channel coefficients
$\{ h_{\ell,m} \}$ in (\ref{hlm}). Each $h_{\ell,m}$ consists of
the sum of gains of all paths whose delay and Doppler shifts lie
within the $(\ell,m)$-th delay-Doppler resolution bin
$S_{\tau,\ell} \cap S_{\nu,m}$ of size $\Delta \tau \times \Delta
\nu$, $\Delta \tau = \frac{1}{W}, \Delta \nu = \frac{1}{T}$ as
illustrated in Fig.~\ref{fig:del_dopp}(a). Distinct $h_{\ell,m}$'s
correspond to approximately \emph{disjoint} subsets of paths and
are hence approximately statistically independent. In this work,
we assume that the channel coefficients $\{ h_{\ell,m}\}$ are
perfectly independent. We also assume\footnote{Note that the Rayleigh 
fading assumption is used only for mathematical tractability. The general 
theme of results will continue to hold as long as the fading distributions 
have an exponential tail. See~\cite{borade} for details 
and~\cite{capacity_sparse_jstsp} 
for a discussion on modeling issues.} Rayleigh fading in which $\{ h_{\ell,m}\}$ 
are zero-mean Gaussian random variables. 

Let $D$ denote the number of non-zero channel
coefficients that reflects the (dominant) statistically independent
DoF in the channel and also signifies the
delay-Doppler diversity afforded by the
channel~\cite{akbar_behnaam}. We decompose $D$ as $D  = D_T D_W$
where $D_T$ denotes the Doppler$/$time diversity and $D_W$ denotes the
frequency$/$delay diversity. The channel DoF or delay-Doppler
diversity is bounded as 
\begin{eqnarray}
D  & = &  D_{T} D_{W} \hspp 
\leq \hspp D_{\max} \triangleq \Dtmax \Dwmax 
\\
\Dtmax & = & \left \lceil TW_d  \right \rceil \ , \ \Dwmax =
\left\lceil T_m W \right \rceil \label{del_dopp_div}
\end{eqnarray}
where $\Dtmax$ denotes the maximum Doppler diversity and $\Dwmax$
denotes the maximum delay diversity. Note that $\Dtmax$ and $\Dwmax$
increase linearly with $T$ and $W$, respectively, and thus represent a
\emph{rich multipath} environment in which each resolution bin in
Fig.~\ref{fig:del_dopp}(a) corresponds to a dominant channel
coefficient.

However, there is growing experimental
evidence~\cite{molisch,saadane,chiachin_residential,porrat2,karedal,molisch_etal,chiachin1} 
that the dominant
channel coefficients get sparser in delay as the bandwidth
increases. Furthermore, we are also interested in modeling
scenarios with Doppler effects, due to motion. In such cases, as
we consider large bandwidths and$/$or long signaling durations, the
resolution of paths in both delay and Doppler domains gets finer,
leading to the scenario in Fig.~\ref{fig:del_dopp}(a) where the
delay-Doppler resolution bins are sparsely populated with paths,
i.e.\ $D \ll D_{\max}$.

In this work, we 
model multipath sparsity by a {\em sub-linear scaling} of 
$D_T$ and $D_W$ with $T$ and $W$, respectively:
\begin{equation}
D_W \sim  \gone(W) \ , \ D_T \sim \gtwo(T) \label{sparse}
\end{equation}
where $\gone$ and $\gtwo$ are \emph{arbitrary} sub-linear
functions. As a concrete example, we will focus on a 
power-law scaling for the rest of this paper:
\begin{equation}
D_{T} = \left( T W_d \right)^{\delta_1} , \hsp \hsp \hsp
D_{W} = \left( W T_m \right)^{\delta_2}  \label{D_powlaw}
\end{equation}
for some $\delta_1,\delta_2 \in (0,1)$. But the results derived here
hold true for any general sub-linear scaling law. Note that
(\ref{del_dopp_div}) 
and (\ref{sparse}) imply that in sparse multipath, 
the total number of delay-Doppler DoF, $D = D_T D_W$, scales
\emph{sub-linearly} with the signal space dimension $N = TW$. 

\begin{rem}
\label{rem1} With perfect CSI at the receiver, the parameter $D$
denotes the delay-Doppler diversity afforded by the channel,
whereas with no CSI, it reflects the level of channel uncertainty;
the number of channel parameters that need to be learned at the
receiver for coherent processing.
\end{rem}

\subsection{Orthogonal Short-Time Fourier Signaling}
\label{sec2b} We consider signaling using an orthonormal
short-time Fourier (STF) basis~\cite{ke_tamer_say,kozek} that is a
natural generalization\footnote{STF signaling can be treated as OFDM
signaling over a block of OFDM symbol periods with an appropriately
chosen symbol duration.} of orthogonal frequency-division
multiplexing (OFDM) for time-varying channels. An orthogonal STF basis
$\{ \phi_{\ell m}(t) \}$ for
the signal space is generated from a fixed prototype waveform $g(t)$ via
time and frequency shifts: $\phi_{\ell m}(t) = g(t-\ell T_o)e^{j2\pi
W_ot}$, where $T_o W_o = 1$, $\ell = 0,\cdots,N_{T}-1$, 
$m=0,\cdots,N_{W}-1$ and $N = N_{T}N_{W} = TW$ with $N_{T} =
T/T_{o}, N_{W}=W/W_{o}$. The transmitted signal can be represented
as
\begin{equation}
x(t) = \sum \limits_{\ell=0}^{N_{T}-1} \sum
\limits_{m=0}^{N_{W}-1} x_{\ell m} \phi_{\ell m}(t) \hsp \hsp \,\
\hsp 0 \leq t \leq T \label{stf_mod}
\end{equation}
where $\{ x_{\ell m} \}$ denote the $N$ transmitted symbols
that are modulated onto the STF basis waveforms. The received
signal is projected onto the STF basis waveforms to yield
\begin{equation}
y_{\ell m} = \langle y, \phi_{\ell m} \rangle = \sum
\limits_{\ell^{'},m^{'}} h_{\ell m, \ell^{'} m^{'}} \;\
x_{\ell^{'} m^{'}} + w_{\ell m}. \label{stf_rx}
\end{equation}
We can represent the system using an $N$-dimensional matrix
equation~\cite{ke_tamer_say,kozek} 
\begin{equation}
\by  = \hsp {\mathbf{H}} \bx + \bw \label{disc_channel}
\end{equation}
where $\bw$ is the additive noise vector whose entries are
i.i.d.\ ${\cal CN}(0,1)$. The $N \times N$ matrix $\bH$ consists of the
channel coefficients $\{h_{\ell m, \ell^{'} m^{'}}\}$ in
(\ref{stf_rx}). We assume that the input symbols that form the
transmit codeword $\bx$ satisfy an average power constraint
\begin{equation}
\frac{1}{T} \cdot \bEe \left[ \| \bx \|^{2} \right] \leq P. 
\label{avg_const}
\end{equation}
Since there are $N=TW$ symbols per codeword, we define the
parameter $\snr$ (transmit energy per modulated symbol) for a
given average transmit power $P$ as $ \snr = \frac{TP}{TW} =
\frac{P}{W}$. In this work, the focus is on the wideband regime
where $\snr \rightarrow 0$ as $W \rightarrow \infty$ for a fixed
$P$.

For sufficiently underspread channels, the parameters $T_o$ and
$W_o$ can be matched to $T_m$ and $W_d$ so that the STF basis
waveforms serve as approximate eigenfunctions of the channel
\cite{kozek,ke_tamer_say}; that is, (\ref{stf_rx}) simplifies
to\footnote{The STF channel coefficients are different from the
delay-Doppler coefficients, even though we are reusing the same
symbols.} $y_{\ell m} \approx h_{\ell m} x_{\ell m} + w_{\ell m}$.
Thus the channel matrix $\bH$ is approximately diagonal. In this
work, we assume that $\bH$ is exactly diagonal; that is,
\begin{equation}
{\mathbf{H}}  =  {\mathrm{diag}} \Big[ \underbrace{ {h}_{1 1}
\cdots {h}_{1 \Ncoh}}_{  {\mathrm{Subspace}} \hsp 1}, \hsp
\underbrace { {h}_{2 1} \cdots {h}_{2 \Ncoh} }_{
{\mathrm{Subspace}} \hsp 2} \hsp  \cdots \hsp \underbrace { {h}_{D
1} \cdots {h}_{D \Ncoh} }_{ {\mathrm{Subspace}} \hsp D} \Big].
\label{H_diag}
\end{equation}
The diagonal entries of $\bH$ in (\ref{H_diag}) admit an intuitive
block fading interpretation in terms of {\em time-frequency
coherence subspaces} \cite{ke_tamer_say} illustrated in
Fig.~\ref{fig:del_dopp}(b). The signal space is partitioned as $N
= TW=\nc D$ where $D$ represents the number of statistically
independent time-frequency coherence subspaces, reflecting the DoF
in the channel, and $\nc$ represents the dimension of each
coherence subspace, which we refer to as the {\bf \em coherence
dimension}. In the block fading model in (\ref{H_diag}), the
channel coefficients over the $i$-th coherence subspace $h_{i 1},
\cdots, h_{i \Ncoh}$ are assumed to be identical (denoted by
$h_i$), whereas the coefficients across different coherence
subspaces are independent and identically distributed. Thus, the
channel is characterized by the $D$ distinct STF channel
coefficients, $\{ h_i \}$, that are i.i.d. zero-mean Gaussian
random variables (Rayleigh fading) with (normalized) variance
equal to $\bEe[|h_i|^2]= \sum_n \bEe[|\beta_n|^2] = 1$
\cite{ke_tamer_say}.

Using the DoF scaling for sparse channels in (\ref{sparse}), the
scaling behavior for the coherence dimension can be computed as
\begin{eqnarray}
\Wcoh & = &  \frac{W}{D_W} \sim \fone(W), \hsp \hsp \hsp \Tcoh =
\frac{T}{D_T} \sim \ftwo(T) 
\\ \nc & = & \Wcoh \Tcoh \sim \fone(W) \ftwo(T) \label{Ncoh}
\end{eqnarray}
where $\Tcoh$ is  the {\em coherence time} and $\Wcoh$ is the {\em
coherence bandwidth} of the channel, as illustrated in
Fig.~\ref{fig:del_dopp}(b). As a consequence of the sub-linearity
of $\gone$ and $\gtwo$ in (\ref{sparse}), $\fone$ and $\ftwo$ are
also sub-linear. In particular, corresponding to the power-law
scaling in (\ref{D_powlaw}), we obtain
\begin{equation}
\Tcoh = \frac{T^{1-\delta_1}}{W_{d}^{\delta_1}}, \hsp \hsp \hsp
\Wcoh = \frac{W^{1-\delta_2}}{T_{m}^{\delta_2}}. \label{nc_powlaw}
\end{equation}
\begin{rem}
\label{rem2} Note that when the channel is sparse, both $\nc$ and
$D$ increase sub-linearly with $N$, whereas when the channel is
rich, $D$ scales linearly with $N$, while $\nc$ is fixed.
\end{rem}

In this work, the focus is on computing achievable rates in the non-coherent 
setting with feedback and as we will see in
Sec.~\ref{sec3} and~\ref{sec4}, the rates turn out to be a
function only of the parameters $\nc$ and $\snr$. Thus, in order
to analyze the low-$\snr$ asymptotics, the following relation
between $\nc$ and $\snr$ $(= P/W)$ plays a key role:
\begin{equation}
\nc  =  \frac{1}{\snr^{\mu}} \hsp , \hsp \hsp \hsp \mu > 0
\label{nc_snr_relation}
\end{equation}
where the parameter $\mu$ reflects the level of channel coherence.
We will revisit (\ref{nc_snr_relation}) and discuss its
achievability and implications in Sec.~\ref{sec4}.

\section{Achievable Rates with Perfect Receiver CSI and Limited 
Channel State Feedback}
\label{sec3} In this section, we study the scenario when there is
perfect CSI at the receiver. We assume throughout this paper that
both the transmitter and the receiver have statistical CSI -
knowledge of $T_m$, $W_d$, $\gone$, $\gtwo$, $\fone$ and $\ftwo$
so that the scaling in $D$ and $\nc$ are known. On one extreme,
with perfect receiver CSI and no transmitter CSI (no feedback),
the coherent capacity per dimension (in nats/s/Hz) equals
\begin{equation}
C_{\coh,0}(\snr) = \sup \limits_{ {\bQ} \hsppp : \hsppp 
{\mathrm{Tr}}({\bQ})
\hspp \leq \hspp TP} \frac{{\bEe} \left[ \log \det \left( \bI_{\nc
D} + {\bH} {\bQ} {\bH}^H \right) \right] }{\nc D}. 
\label{coh_cap_rxcsi}
\end{equation}
The optimization is over the set of $\nc D$-dimensional positive
definite input covariance matrices $\bQ = \bEe \left[ \bx \bx^{H}
\right]$ satisfying the average power constraint in
(\ref{avg_const}). Due to the diagonal nature of $\bH$ in
(\ref{H_diag}), the optimal $\bQ$ is also diagonal. Furthermore,
with no transmitter CSI, the uniform power allocation $\bQ =
\frac{TP}{\nc D} \hspp \bI_{\nc D} = \snr \cdot \bI_{\nc D}$
achieves this optimum. The corresponding capacity in the limit of
low-$\snr$ is~\cite{verdu,zheng} 
\begin{equation}
C_{\coh,0} (\snr) \approx 
\snr - \snr^{2}. 
\label{coh_cap_lowsnr}
\end{equation}

On the other extreme is the case of perfect receiver and
transmitter CSI, where the receiver instantaneously feeds back all
the channel coefficients, $\{ h_i \}_{i=1}^{D}$, 
corresponding to the $D$ independent coherence
subspaces to the transmitter. The optimum transmitter power allocation in this case
is waterfilling~\cite{gallager_book,goldsmith_varaiya} over the different
coherence subspaces. In the low-$\snr$ extreme, it is shown in
\cite{verdu,borade} that the capacity with perfect transmitter CSI
scales as $\log \left(\frac{1}{\snr} \right) \snr$. That is, the
capacity gain (compared with the receiver CSI only case) is directly
proportional to the waterfilling threshold, $h_{w} \sim \log
\left(\frac{1}{\snr} \right)$, and this gain serves as a benchmark
for all limited feedback schemes. More interestingly, it is shown
in \cite{verdu,borade} that this maximum capacity gain can be achieved
with just one bit of feedback per channel coefficient.

In the case of limited feedback, both the transmitter and the
receiver have \emph{a priori} knowledge of a common threshold denoted by 
$\HT$. The receiver compares the channel strength ($|h_i|^{2},\hsp
i=1,2,\cdots,D$) in each coherence subspace with $\HT$, and feeds
back
\begin{equation}
b_i= \begin{cases} 1  \hsp \hsp {\rm if} \hsp \hsp |h_i|^{2} \geq \HT \\
0 \hsp \hsp {\rm if} \hsp \hsp |h_i|^{2} < \HT. \end{cases}
\label{one_bit_fb}
\end{equation}
At the transmitter, power allocation is uniform across the coherence subspaces 
for which $b_i = 1$ and no power is allocated to those subspaces for which 
$b_i = 0$. The input power allocation is conditioned on the partial CSI 
available at the transmitter (denoted by ${\sf CSI}$), which is $\{ b_i \}_{i=1}^{D}$. 
This power allocation, which we still denote by $\bQ$ with an abuse of notation, 
takes the form 
\begin{eqnarray}
\bQ ({\sf CSI}) 
& = & \diag \left( \bEe[  |x_{1}|^{2} | {\sf CSI} ] , \bEe[ |x_{2}|^{2} | {\sf CSI}] 
, \cdots,
\bEe [ |x_{N}|^{2} | {\sf CSI}] \right)  
\\
& = & \diag \big (\underbrace{ q_1,\cdots,q_1 }_{\nc} \hspp ,
\hspp \underbrace{ q_2,\cdots,q_2 }_{\nc} \hspp ,\cdots, \hspp
\underbrace{ q_D,\cdots,q_D}_{\nc} \big) 
\\ q_i & = & \po \cdot \chi( |h_i|^2 \geq \HT ). \label{Q_txcsi}
\end{eqnarray}
The choice of $\po$ depends on the type of power constraint and
also on the nature of feedback. To explore this further, let $\De$
denote the number of {\em active} subspaces, those which exceed the
threshold $\HT$. We have
\begin{equation}
\De =  \sum \limits_{i=1}^{D} \chi( |h_i|^2 \geq \HT )
\label{Deff}
\end{equation}
\begin{equation}
\bEe \left[ \De \right] \stackrel{(a)}{=} D \bEe \left[
\chi( |h|^2 \geq \HT ) \right] \stackrel{(b)}{=} D e^{-\HT}
\label{EDeff}
\end{equation}
where (a) is due to the fact that $\{ h_i \}_{i=1}^{D}$ are i.i.d.
and (b) is due to the fact that for a standard Gaussian, $\bEe \left[
\chi( | h_i |^2 \geq \HT ) \right] = \prob \left( | h_i |^2 \geq
\HT \right) = e^{-\HT}$.

If we assume knowledge of $\{ b_i \}_{i=1}^{D}$ at the 
{\em beginning} of each codeword, albeit non-causally, 
at the transmitter, then we can uniformly divide power
among the active subspaces. That is
\begin{equation}
\ponc = \frac{TP}{\nc \De}. \label{po_nc}
\end{equation} 
The rate achievable with this power allocation, denoted by $C_{\coh,1,{\sf LT}}(\snr)$, 
is 
\begin{equation}
C_{\coh,1,{\sf LT}}(\snr) 
=  \max \limits_{\HT}
\frac{1}{D} \sum_{i = 1}^{D} \bEe \left[ \log \left( 1 +
\frac{TP}{\nc \De} \cdot \left| h_i \right|^2 \right) \chi \left(
|h_i|^2 \geq \HT \right) \right ]. \label{cap_coh_nc}
\end{equation}
The power allocation in (\ref{po_nc}) satisfies the power
constraint instantaneously as well as on average. To see this,
note that
\begin{equation}
P_{\inst,\nca} = 
\frac{\nc}{T} \sum \limits_{i=1}^{D} q_i 
=  \frac{\nc}{T} \sum \limits_{i=1}^{D}
\frac{TP}{\nc \De} \chi \left( |h_i|^2 \geq \HT \right) =  P
\label{pinst_nc}
\end{equation}
and clearly $\bEe \left[P_{\inst,\nca} \right] = P$ as well.
The non-causality of the scheme is more relevant in the 
scenario when the receiver estimates the channel coefficients $\{
h_{i} \}_{i=1}^{D}$ and feeds back $\{ b_{i} \}_{i=1}^{D}$ based
on these estimates. This motivates us to instead consider a causal
power allocation scheme, one in which for all $i=1,\cdots,D$,
$q_{i}$ in (\ref{Q_txcsi}) depends on $b_{i}$ only through the indicator 
function and $\po$ is
independent of $\{ b_{i} \}_{i=1}^{D}$. From (\ref{Q_txcsi}), we
have
\begin{equation}
\bEe \left[ \| \bx \|^{2} \right] = \nc \sum \limits_{i=1}^{D}
\bEe \left[ q_i \right] = \nc \sum \limits_{i=1}^{D} \po \cdot
\bEe \left[ \chi( |h_i|^2 \geq \HT ) \right] 
=\nc \po \bEe \left[\De \right]. 
\label{eq_avg_pow}
\end{equation}
Thus to satisfy $\bEe \left[
\| \bx \|^{2} \right] \leq TP$, the power allocation for the 
causal scheme is given by
\begin{equation}
\poc = \frac{TP}{\nc \bEe \left[ \De \right]} = \frac{TP}{\nc D
e^{-\HT}} \label{po_c}
\end{equation}
and the corresponding rate, $\widehat{C}_{\coh,1,{\sf LT}}(\snr)$, is given by
\begin{equation}
\widehat{C}_{\coh,1,{\sf LT}}(\snr) 
= \max \limits_{\HT} \frac{1}{D} \sum_{i=1}^D
\bEe \left[ \log \left( 1 + \frac {TP}{\nc D e^{-\HT} } | h_i
|^2 \right) \chi( \left| h_i \right|^2 \geq \HT ) \right].
\label{cap_coh_c}
\end{equation}
The causal power allocation policy in (\ref{po_c}) satisfies the
average power constraint but can have a large instantaneous power.
This is because 
\begin{equation}
P_{\inst,\ca} = \frac{\nc}{T} \sum \limits_{i=1}^{D} \frac{TP}{\nc
D e^{-\HT} } \chi \left( |h_i|^2 \geq \HT \right) =
\left(\frac{\De}{D e^{-\HT}} \right) P. \label{pinst_c}
\end{equation}
Thus $\bEe \left[ P_{\inst,\ca} \right] = P$, but unlike
(\ref{pinst_nc}), $P_{\inst,\ca} \in [0,\infty)$ depending on the choice of 
$\HT$. We will address
this issue in Sec.~\ref{sec3b}, but first, we study the average power 
constraint case more carefully.

\subsection{Achievable Rates under Average Power Constraint}
\label{sec3a} The following theorem establishes that a threshold
of the form $\HT \sim \lambda \log \left(\frac{1}{\snr} \right)$
for some $\lambda \in (0,1)$ provides the solution to
(\ref{cap_coh_c}).

\begin{thm}
\label{thm1} 
Given any $\lambda \in (0,1)$, a causal on-off signaling scheme under an 
average power constraint achieves
$\widehat{C}_{\sf LB } \leq \widehat{C}_{\coh,1,\LT}(\snr) \leq
\widehat{C}_{\sf UB}$ 
with an optimal threshold of the form: 
\begin{eqnarray}
\label{HT_eqn} \lim \limits_{\snr \rightarrow 0}
\frac{\HT}{\lambda \log \left(\frac{1}{\snr} \right)} = 1 
\end{eqnarray}
where 
\begin{eqnarray}
\widehat{C}_{\sf UB} & = & \snr^{\lambda} \cdot \left[\log \left(
\tsty{ 1 + \lambda \snr^{1- \lambda} \log\left( \frac{1} {\snr}
\right) } \right) + \log \left( \tsty{ 1 +
\frac{\snr^{1-\lambda} } { 1 + \lambda \snr^{1-\lambda} \log
\left( \frac{1}{\snr} \right) } } \right) \right] \label{chat_ubd}
\\
\widehat{C}_{\sf LB} & = & \snr^{\lambda} \cdot \left[\log \left(
\tsty{ 1 + \lambda \snr^{1- \lambda} \log\left( \frac{1} {\snr}
\right) } \right) + \frac{1}{2} \log \left( \tsty{ 1 + \frac{2
\snr^{1-\lambda} } { 1 + \lambda \snr^{1-\lambda} \log \left(
\frac{1}{\snr} \right) } } \right) \right]. \label{chat_lbd}
\end{eqnarray}
\end{thm}
\vspace{0.2in}
\noindent 
\begin{proof}
Starting from (\ref{cap_coh_c}), we have 
\begin{eqnarray}
\widehat{C}_{\coh,1,\LT}(\snr) 
& =  &\max \limits_{\HT} \frac{1}{D} \sum_{i=1}^D
\bEe \left[ \log \left( 1 + \frac {TP}{\nc D e^{-\HT} } | h_i |^2 \right) 
\chi( \left| h_i \right|^2 \geq \HT ) \right] 
\\ 
& \stackrel{(a)}{=} & \bEe \left[ \log \left( 1 +
\snr \hspp e^{\HT} \hspp  | h |^2 \right) \hspp \chi( \left| h
\right|^2 \geq \HT ) \right] \label{cap_expec}
\end{eqnarray}
where (a) follows from the fact that $\{ h_i \}$ are i.i.d.\ ${\cal CN} (0,1)$ 
and $h$ is a generic i.i.d.\ ${\cal CN}(0,1)$ random variable. 
The expectation in (\ref{cap_expec}) can be computed using
\cite[4.337(1), p.\ 574]{gradshteyn}. With 
$\alpha \triangleq \frac{1 + \snr \hspp \HT \hspp e^{\HT} 
}  { \snr \hspp e^{\HT} }$, we have
\begin{eqnarray}
\widehat{C}_{\coh,1,\LT}(\snr) 
 & = & e^{- \HT} \cdot \left[ \tsty{ \log \left(1
+ \snr \hspp \HT \hspp e^{\HT} \right) } + \exp
\left( \alpha \right) \int_{ \alpha }^{\infty} \frac{ e^{-t} }{t} \hspp 
\ud t \right] 
\\ 
& = & e^{- \HT} \cdot \left[ \tsty{ \log \left(1 +
\snr \hspp \HT \hspp e^{ \HT} \right) } + 
\nu_{\alpha} \right] \label{cap_widehat_eqn}
\end{eqnarray}
where 
$\nu_{\alpha} \triangleq 
\exp \left( \alpha \right) \int_{ \alpha }^{\infty}
\frac{ e^{-t} }{t} \hspp \ud t.$ 
As $\alpha \rightarrow \infty$, 
the following bounds hold for $\nu_{\alpha}$~\cite[5.1.20, p.\ 229]{stegun}:
\begin{equation}
\frac{1}{2} \log \left(1 + \frac{2}{\alpha} \right) \leq
\nu_{\alpha} \leq \log \left(1 +\frac{1}{\alpha} \right).
\label{nu_bounds}
\end{equation}
It can be checked that the choice of $\HT$ maximizing (\ref{cap_widehat_eqn}) 
is obtained by setting its derivative to zero and satisfies
\begin{eqnarray}
\Delta \triangleq 1 - \log \left( 1 + \snr \hspp \HT e^{\HT} \right) -
\frac{1}{ \snr e^{\HT} } \cdot \nu_{\alpha} = 0.
\label{delta_zero}
\end{eqnarray}
Now, if $\HT$ is such that $ \lim \limits_{ \snr \rightarrow 0} \frac{\HT}{
\lambda \log \left( \frac{1}{\snr} \right) } = 1$ for some
$\lambda \in (0,1)$, then as $\snr \rightarrow 0$, we have $\snr
\hspp \HT \hspp e^{\HT} \rightarrow 0$ and $\alpha \rightarrow
\infty$. Thus using (\ref{nu_bounds}), we can approximate 
$\nu_{\alpha}$ as $\nu_{\alpha} \approx
\frac{1}{\alpha}$. With this approximation in (\ref{delta_zero}), we have
$\frac{1}{ \snr e^{\HT} } \cdot \nu_{\alpha} \approx \frac{1}{1+
\snr \HT e^{\HT}} \rightarrow 1$. Using the choice of
$\HT$ as in (\ref{HT_eqn}), it follows that as $\snr \rightarrow
0$, $\Delta \rightarrow 0$. Substituting this choice of $\HT$ in
(\ref{cap_widehat_eqn}) and using the upper and lower bounds on
$\nu_{\alpha}$ in (\ref{nu_bounds}), we obtain the bounds in
(\ref{chat_ubd}) and (\ref{chat_lbd}).
\end{proof}

It can also be shown that the rate achievable with the causal
scheme is asymptotically (in low-$\snr$) the same as the
non-causal capacity in (\ref{cap_coh_nc}). That is,
$\widehat{C}_{\coh,1,\LT}(\snr)$ is a tight bound to
$C_{\coh,1,\LT}(\snr)$ and for all $\lambda \in (0,1)$, we have
\begin{eqnarray}
\lim_{\snr \rightarrow 0} \frac{ \big| C_{\coh,1,\LT}(\snr) -
\widehat{C}_{\coh,1,\LT}(\snr) \big|  } { C_{\coh,1,\LT}(\snr) } &
= & 0. \label{chat_approx}
\end{eqnarray}
The proof of the above statement can be found in
Appendix~\ref{app_closeness}.

\begin{cor}
\label{cor1} The capacity gain for the $D$-bit channel state feedback, causal
power allocation scheme over the capacity with only receiver CSI in 
(\ref{coh_cap_lowsnr}) is 
\begin{equation}
\lim_{\snr \rightarrow 0} 
\frac{ \widehat{C}_{\coh,1,\LT}(\snr) }{ C_{\coh,0}(\snr)} = (1+
\HT) =  1 + \lambda \log \left(\frac{1}{\snr} \right).
\end{equation}
\end{cor} 
{\vspace{0.1in}}
\noindent \begin{proof}
A Taylor series expansion of the upper and lower
bounds in (\ref{chat_ubd}) and (\ref{chat_lbd}) shows that they
are equal up to first-order. This common term is such that 
\begin{eqnarray}
\widehat{C}_{\coh,1,\LT}(\snr) = \snr \hspp 
\left(  1+\lambda \log \left(
\frac{1}{\snr} \right) \right) = (1+\HT) \snr. 
\end{eqnarray}
On the other hand, with CSI at the receiver alone, we have from (\ref{coh_cap_lowsnr}),
$\frac{C_{\coh,0}(\snr) }{\snr} = \left(1 + {\rm o}(1) \right)$. 
Thus the desired result follows.
\end{proof}
\begin{rem}
\label{rem3} The capacity gain due to feedback is directly
proportional to $\HT$ and the highest gain is obtained by choosing
$\lambda \rightarrow 1$, and equals the benchmark where perfect CSI is available 
at both the ends~\cite{borade}. Statements analogous to those in 
Theorem~\ref{thm1} and Corollary~\ref{cor1} are well-known from prior work; 
see~\cite{verdu,borade,manish} for details. 
\end{rem}

We now revert our attention back to the instantaneous transmit
power case described in (\ref{pinst_c}). Note that as $D \rightarrow
\infty$, $P_{\inst,\ca} \rightarrow P$ as a consequence of the law
of large numbers. However, for any finite $D$,
$P_{\inst,\ca}$ may be much larger than $P$. This is a serious
issue in practical systems that typically operate with peak power
limitations. Thus it is important to analyze the impact of
constraints on the instantaneous power in (\ref{pinst_c}), as
discussed next.

\subsection{Achievable Rates under Instantaneous Power Constraint}
\label{sec3b} In addition to the average power constraint, let us 
impose a constraint on the instantaneous transmit power of the
form
\begin{equation}
P_{\inst,\ca} \stackrel{a.s.}{\leq} A P \label{ST_const}
\end{equation}
where $A > 1$ is finite. With this short-term constraint, we now compute 
the rate, $\widehat{C}_{\coh,1,\ST}(\snr)$, achievable with the 
causal signaling scheme. We are particularly interested in
exploring conditions under which 
$\widehat{C}_{\coh,1,\ST}(\snr) \approx
\widehat{C}_{\coh,1,\LT}(\snr)$. To this end, we employ the
following power allocation
\begin{eqnarray}
\bQ  & = & \diag \big (\underbrace{ q_1,\cdots,q_1 }_{\nc} \hspp ,
\hspp \underbrace{ q_2,\cdots,q_2 }_{\nc} \hspp ,\cdots, \hspp
\underbrace{ q_D,\cdots,q_D}_{\nc} \big) \\
q_i  & = & \poc \: \chi( |h_i|^2 \geq \HT ) \: \chi \left( \tsty{ \sum
\limits_{j = 1}^i \chi ( |h_j|^2 \geq \HT ) \leq A D e^{-\HT} }
\right). \label{Q_inst}
\end{eqnarray}
The second indicator function in (\ref{Q_inst}) checks for the
constraint in (\ref{ST_const}) causally, during each
time-frequency coherence slot, and allocates power only if this
constraint is met. Note that the choice of $q_i$ in~(\ref{Q_inst}) meets the 
average power constraint with an inequality and hence, $q_i$ can be 
enhanced further. On the other hand, the right-hand side of the argument 
within the second indicator function has to be reduced by the factor 
$\frac{T_i}{T}$ where $T_i$ corresponds to the time duration over which 
the $i$ coherence subspaces under consideration are encountered. We will not 
bother with these secondary issues in the ensuing analysis. We then have 
\begin{eqnarray}
\begin{split}
\widehat{C}_{\coh,1,\ST}(\snr) \\
& {\hspace{-0.8in}} = \frac{1}{D} \hspp \bEe \left[ \sum_{i=1}^D
\log \left( 1 + \frac{TP}{\nc} |h_i|^2 \frac{ \chi ( |h_i|^2
\geq \HT ) }{D e^{-\HT}} \chi \left( \sum_{j = 1}^i \chi ( |h_j|^2
\geq \HT ) \leq A D e^{-\HT} \right)
 \right) \right] \label{cap_ST} \\
& {\hspace{-0.8in}} =  \frac{1}{D} \sum_{i=1}^D \bEe \left[
\log \left( 1 + \snr \cdot e^{\HT} \cdot |h_i|^2 \chi (
|h_i|^2 \geq \HT ) \right) \chi \left( \sum_{j = 1}^i \chi (
|h_j|^2 \geq \HT )
\leq A D e^{-\HT} \right) \right] \label{eq1} \\
& {\hspace{-0.8in}} = \frac{1}{D} \hspp \sum_{i=1}^D \prob \left(
\sum_{j = 1}^i \chi ( |h_j|^2 \geq \HT ) \leq A D e^{-\HT} \right)
\cdot \bEe \left[ \log \left( 1 + \snr \cdot e^{\HT} \cdot
|h_i|^2 \chi ( |h_i|^2 \geq \HT ) \right)
\right] \label{eq2} \nonumber \\
& {\hspace{-0.8in}} \stackrel{(a)}{=} \bEe \left[ \log \left(
1 + \snr \cdot e^{\HT} \cdot |h|^2 \chi ( |h|^2 \geq \HT ) \right)
\right] \cdot \frac{  \sum_{i=1}^D \prob \left( \sum_{j = 1}^i
\chi ( |h_j|^2 \geq \HT ) \leq A D e^{-\HT} \right) }{D} \\ 
& {\hspace{-0.8in}} 
= \widehat{C}_{\coh,1,\LT}(\snr) \cdot \frac{\sum_{i=1}^D p_i}{D}
\label{cap_ST_eqn}
\end{split}
\end{eqnarray}
where $\widehat{C}_{\coh,1,\LT}(\snr)$ is the rate achievable with only an average
power constraint, and (a) follows from the fact that $\{h_i \}$
are i.i.d.\ and 
\begin{eqnarray} 
p_i \triangleq \prob \left( \sum_{j = 1}^i \chi (
|h_j|^2 \geq \HT ) \leq A D e^{-\HT} \right). 
\end{eqnarray}
Thus,
characterizing $\widehat{C}_{\coh,1,\ST}(\snr)$ is equivalent to
computing $p_i$. In particular, under what condition does
$\frac{\sum_{i=1}^D p_i}{D} \rightarrow 1$? This is discussed in
the following proposition.
\begin{prp}
\label{prop_peak} With $\HT \sim \lambda \log \left(
\frac{1}{\snr} \right)$ as in~(\ref{HT_eqn}), we have 
$\frac{\sum_{i=1}^D p_i}{D} \geq L$ where 
\begin{equation}
L \approx 1 - \tsty{ \frac{4}{\snr^{\lambda} \left(1 +
\snr^{\lambda}/4 \right) ^{\frac{AD}{2} -1 } } - \frac{D(1 - A/2)}
{\left( 1 + \snr^{\lambda}/4 \right)^{D(A-1)^2}} } \label{L1}
\end{equation} 
if $1 < A < 2$, and if $A > 2$, we have
\begin{equation}
L \approx 1 - \frac{4} {\tsty{ \snr^{\lambda}} \left( 1 +
\tsty{\snr^{\lambda}/4} \right)^{D(A-1)} }. \label{L2}
\end{equation}
In particular, if
\begin{equation}
\bEe \left[ \De \right] - \HT = D \hspp e^{-\HT} - \HT 
\sim  D \snr^{\lambda} + \lambda \log(\snr) 
\rightarrow \infty \hsp {\rm as} \hsp \snr \rightarrow 0, 
\label{ST_cond}
\end{equation}
then $L \rightarrow 1$ for all $A > 1$ and
$\widehat{C}_{\coh,1,\ST}(\snr) \rightarrow
\widehat{C}_{\coh,1,\LT}(\snr)$ .
\end{prp}
{\vspace{0.1in}}
\begin{proof}
See Appendix~\ref{app_ltpc}.
\end{proof}

\subsection{Discussion: Rich vs. Sparse Multipath}
\label{sec3c} The result of Theorem~\ref{thm1} implies that the rate 
achievable with the $D$-bit channel state 
feedback scheme approaches the benchmark, the perfect transmitter CSI 
capacity when $\lambda \rightarrow 1$. 
Furthermore, this benchmark can be attained in
the wideband limit, \emph{even} when there is an instantaneous
power constraint. As described in Prop.~\ref{prop_peak},
$\bEe \left[\De \right] - \HT \rightarrow \infty$ provides a sufficient
condition. We now discuss the feasibility of satisfying these
conditions when the channel is rich and when it is sparse. 
The behavior of $\bEe \left[\De \right]$ 
provides key insights in this regard. 

\noindent \textbf{A1)} {\textbf{Rich multipath}}: For a rich
channel, from (\ref{del_dopp_div}) we note that $D$ scales
linearly with $T$ and $W$. For a fixed $T$, $D \sim
\snr^{-1}$ (since $\snr = \frac{P}{W}$). That is, $\bEe \left[\De
\right] - \HT = D \: \snr^{\lambda} + \lambda \log(\snr) 
\rightarrow \infty$ for $0 < \lambda < 1$. We can thus conclude that for 
rich multipath the perfect CSI benchmark is attained trivially with both 
average and instantaneous power constraints.

\noindent \textbf{A2)} {\textbf{Sparse multipath}}: From the
power-law scaling in (\ref{D_powlaw}), ignoring the 
constant factors, we have $D \sim T^{\delta_1} W^{\delta_2}$ and therefore
\begin{equation}
\bEe \left[\De \right] - \HT 
\sim T^{\delta_1} \snr^{\lambda - \delta_2} + \lambda 
\log(\snr). \label{EDeff_fixT}
\end{equation}
For a fixed $T$, as $\snr \rightarrow 0$, we have 
\begin{equation}
\bEe \left[\De \right] - \HT 
\rightarrow \begin{cases} 
\infty & {\rm if} \hsp  0 < \lambda < \delta_2  \\ 
-\infty & 1 > {\rm if} \hsp \lambda \geq \delta_2. 
\end{cases} \label{Deff_fixT}
\end{equation}
While we can approach the benchmark capacity with an average power
constraint, (\ref{Deff_fixT}) suggests a cap on $\lambda$, 
the highest achievable gain with an instantaneous power constraint.


\subsection{Capacity Optimal Packet Configurations}
\label{sec_fb_cap_opt_pkt}  From (\ref{Deff_fixT}), we see 
that the perfect CSI gain is not always achievable
when there is an instantaneous power constraint. However, we note
that (\ref{Deff_fixT}) is derived assuming a \emph{fixed} choice of $T$, 
while we know
that sparsity in Doppler facilitates any desired scaling in the
DoF with increasing $T$.
Leveraging both delay and Doppler sparsities, we propose the
following solution to get around the restriction in
{\bf A2}. Instead of signaling with a fixed duration $T$, let us 
suppose that we maintain a scaling relationship for $T$ as a function
of $W$. For example, let $T \sim W^{\rho}$ for some $\rho
> 0$. Consequently, $D \sim T^{\delta_1} W^{\delta_2} \sim
W^{\delta_2 + \rho \delta_1}$ and we have
\begin{equation}
\bEe \left[\De \right] - \HT \sim \snr^{\lambda
-\delta_2 - \rho \delta_1} + \lambda \log(\snr). 
\label{EDeff_varyT}
\end{equation}
Thus in the limit as $\snr \rightarrow 0$, the asymptotic behavior of 
$\bEe \left[\De \right] - \HT$ is given by 
\begin{equation}
\bEe \left[\De \right] - \HT 
\rightarrow \begin{cases} 
\infty & {\rm if} \hsp 0 < \lambda < \delta_2 + \rho \delta_1 \\ 
-\infty & 1 > {\rm if} \hsp \lambda \geq \delta_2 + \rho \delta_1.  
\end{cases}
\label{Deff_varyT}
\end{equation}
Note that in (\ref{Deff_varyT}), we have
\begin{equation}
\delta_2 + \rho \delta_1 \geq 1 \: \Longleftrightarrow \: \rho \geq
\frac{1-\delta_2}{\delta_1} \label{rho_condition}
\end{equation}
which consequently leads to the desired result that $\bEe
\left[\De \right] - \HT \rightarrow \infty$ for all $\lambda \in (0,1)$.
Thus the benchmark gain is achievable even under an
instantaneous power constraint.

\begin{figure*}
\begin{center}
\centerline{\includegraphics[width=4.0in]{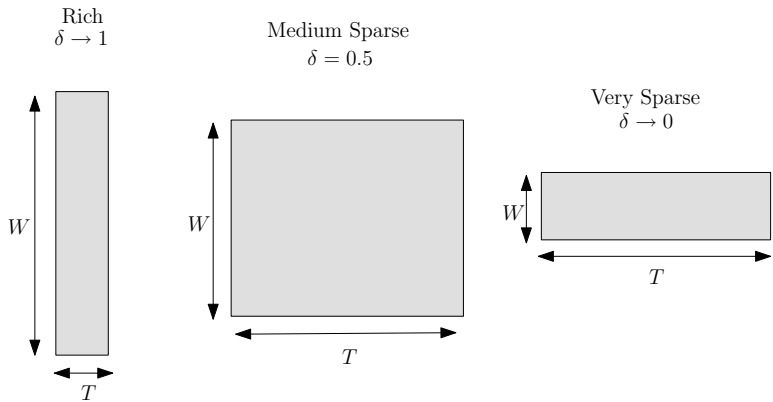}}
\end{center}
\caption{ \sl Optimal packet configurations with perfect receiver
CSI and limited feedback as a function of richness of the channel. 
Three cases are illustrated here: Rich
multipath $(\delta \rightarrow 1)$, medium sparsity $(\delta =
0.5)$ and very high sparsity $(\delta \rightarrow 0)$.}
\label{fig_fb_cap_opt_pkt_coh}
{\vspace{-5mm}}
\end{figure*}

To further illustrate this idea, we present an example when
channel sparsity follows the power-law scaling in
(\ref{D_powlaw}). For simplicity, let us assume that $\delta_1 =
\delta_2 = \delta$. From~(\ref{rho_condition}), 
we require $ T \sim W^{\rho}$ with $\rho \geq \frac{1-\delta}{\delta}$ 
to achieve the benchmark performance. With $N = TW$, the capacity optimal 
$(T,W)$ packet configuration is then given by
\begin{equation}
T \sim N^{\frac{\rho}{1+\rho}}, \hsp \hsp \hsp \hsp W \sim
N^{\frac{1}{1+\rho}}. \label{fb_cap_opt_pkt_coh}
\end{equation}
Fig.~\ref{fig_fb_cap_opt_pkt_coh} illustrates the optimal packet
configuration relationship for a rich multipath channel $(\delta \rightarrow
1)$, for a medium sparse channel $(\delta = 0.5)$ and for a very
sparse channel $(\delta \rightarrow 0)$. They show that in sparse multipath channels,
the perfect CSI capacity gain is achievable with limited feedback
under both average and instantaneous constraints on the
transmission power by appropriate signaling strategies. These guidelines can be
easily extended to generic sub-linear scaling laws. 

\section{Achievable Rates with Channel Estimation at the Receiver}
\label{sec4} In contrast to the perfect receiver CSI case, we now consider 
the more realistic case where no CSI is available {\em a priori}. 
We first consider only an average power constraint and show that
the first-order term of the benchmark capacity can be achieved if the
channel is sparse and the channel coherence dimension, $\nc$, scales
with $\snr$ at an appropriate rate, allowing the receiver to learn
the channel reliably. We also show that this is infeasible when
the channel is rich, due to poor channel estimation. 

\begin{figure*}
\begin{center}
\centerline{\includegraphics[width=4.0in]{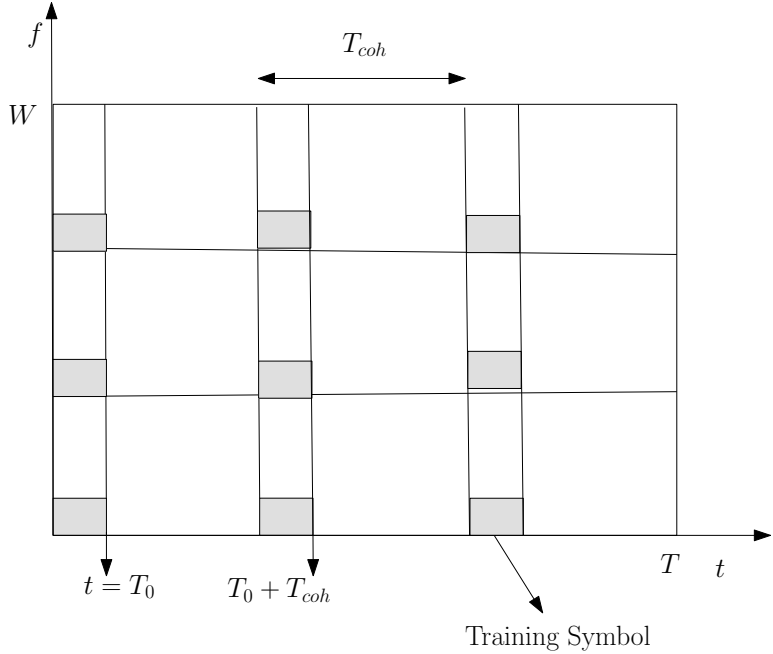}}
\end{center}
\caption{ \sl Training-based signaling scheme in the STF
domain. The $D$ estimated channel coefficients determine the $D$
feedback bits for the communication scheme with limited feedback.}
\label{fig_tr_fback}
\end{figure*}

More specifically, the focus is here on a training-based signaling scheme 
where the transmitted signals
include training symbols to enable channel estimation and coherent
detection. The restriction to training schemes is motivated by
their easy realizability. The total energy available for
training and communication is $PT$, of which a fraction $\eta$ is
used for training and the remaining fraction $(1-\eta)$ is used
in communication. With the block fading model, this means that 
one signal space dimension in each coherence subspace is used for training
and the remaining $\left( \nc-1 \right)$ are used in communication. This is 
pictorially illustrated in Fig.~\ref{fig_tr_fback}. We consider minimum 
mean-squared error (MMSE) channel estimation and the reader
is referred to~\cite[Sec.~IIc]{capacity_sparse_jstsp} for more
details on the training scheme.

{\vspace{-5mm}}

\subsection{Achievable Rates under Average Power Constraint}
\label{sec4b} Let $\widehat{C}_{\train,1,\LT}(\snr)$ denote the
average mutual information achievable (per-dimension) with the
causal training scheme under the average
power constraint. We proceed along the same lines
as the no feedback case~\cite[Lemma 1]{capacity_sparse_jstsp} 
to characterize
$\widehat{C}_{\train,1,\LT}(\snr)$. Let $\bH$ be the actual channel,
$\widehat{\bH}$ be the estimated channel and ${\bf \Delta} = \bH -
\widehat{\bH}$ denote the estimation error matrix. We begin with
the following well-known lower-bound~\cite{medard} to
$\widehat{C}_{\train,1,\LT}(\snr)$:
\begin{eqnarray}
\widehat{C}_{\train,1,\LT}(\snr) 
\geq \sup_{ \bQ } \frac{ {\bEe} \left[ \tsty{
\log \det \tsty{ \big( \bI_{(\nc-1) D} + \widehat{\bH } {\bQ}
\widehat{\bH }^H \left( \bI + \Sigma_{ {\mathbf{\Delta}}
{\mathbf{x}} } \right)^{-1} \big) } } \right] }{\nc D} 
\label{IfbLB}
\end{eqnarray}
where the supremum is over $\{ \bQ : \hspp {\rm Tr}(\bQ) \hspp \leq
\hspp (1-\eta) TP \}$. The optimal $\bQ$ is again diagonal and
analogous to (\ref{Q_txcsi}), equals
\begin{eqnarray}
\bQ & = & \diag \big (\underbrace{ q_1,\cdots,q_1 }_{\nc-1} \hspp
, \hspp \underbrace{ q_2,\cdots,q_2 }_{\nc-1} \hspp ,\cdots, \hspp
\underbrace{ q_D,\cdots,q_D}_{\nc-1} \big)  \\
q_i & = & \frac{(1 - \eta)TP} {(\nc - 1) D} \cdot \frac{ \chi
\left(  | \widehat{h}_i |^2 \geq \HTtr \right) } {\bEe \left[ \chi
\left(  | \widehat{h} |^2 \geq \HTtr \right) \right] } 
\label{Q_trscheme}
\end{eqnarray}
where $\HTtr$ is the threshold in the training case. 
The following theorem describes conditions under which the rates achievable 
with the training scheme converge to those in the coherent case. 

\begin{thm}
\label{thm2} If $\nc = \frac{1}{\snr^{\mu}}$ for some $\mu > 1$,
then
\begin{eqnarray}
\lim_{\snr \rightarrow 0} \frac{ \widehat{C}_{\train,1,\LT}(\snr)
} { \widehat{C}_{\coh,1,\LT}(\snr)} = 1.
\label{desired}
\end{eqnarray}
\end{thm}
{\vspace{0.1in}}
\begin{proof}
Using the choice of $\bQ$ from (\ref{Q_trscheme}) in (\ref{IfbLB})
and proceeding along the lines of (\ref{cap_ST_eqn}), we obtain 
\begin{eqnarray}
\widehat{C}_{\train,1,\LT} \left( \HTtr, \eta, \nc, \snr \right) 
&=& \kappa_1 \cdot \bigg[ \log \left( 1 + \frac{ (1-\eta) (1 +
\eta \nc \snr)
\HTtr \snr } { (1 - \eta) \snr + \kappa_1 \kappa_2} \right)  \nonumber 
\\ 
&& {\hspace{0.25in}} + \nu_{ \frac{ (1 -\eta)(1 + \eta \nc \snr) \HTtr \snr + (1 -
\eta) \snr + \kappa_1 \kappa_2 } {\eta(1 - \eta) \nc \snr^2} }
\bigg], \label{C_tr_simplify} \\
 \kappa_1 & = &  e^{-\frac{ \HTtr (1 + \eta \nc \snr)} {\eta \nc
\snr} }, \: \:
\kappa_2  =  \eta(\nc - 1) \snr + \left(1 - \frac{1}{\nc} \right)
\end{eqnarray}
where $\nu_{\bullet}$ is as defined following (\ref{cap_widehat_eqn}). The tightest
lower bound to (\ref{C_tr_simplify}) is obtained by maximizing
$\widehat{C}_{\train,1,\LT} \left( \HTtr, \eta, \nc, \snr \right)$
over $\eta$, 
the fraction of energy spent on training, and over $\HTtr$:
\begin{eqnarray}{C}^{*}_{\train,1,\LT} & = & \max \limits_{\HTtr} \left[ \max
\limits_{\eta}  \widehat{C}_{\train,1} \left( \HTtr, \eta, \nc,
\snr \right) \right]. \label{max_max}
\end{eqnarray}
Performing the optimization in~(\ref{max_max}) 
seems difficult. Motivated by our study in
Sec.~\ref{sec3}, we now assume a specific form for the
threshold: $\HTtr = \epsilon \log \left(\frac{1}{\snr}
\right)$. It is shown in Appendix~\ref{app_pf_thm2} that with this
choice of $\HTtr$, the optimal choice for $\eta$ and $\nc$
can be obtained in closed form and the desired result in~(\ref{desired}) is
established. 

Alternatively, we 
demonstrate a sub-optimal, but simpler approach that 
suffices to obtain (\ref{desired}). This approach uses the choice of $\eta$ 
that optimizes the average mutual information in the no feedback 
case~\cite[Lemma 2]{capacity_sparse_jstsp}. This choice, denoted by $\etas$, 
is given as 
\begin{eqnarray}
\etas & = & \tsty{ \frac{\nc \snr + \nc - 1}{(\nc-2) \nc
\snr}\cdot \left[ \sqrt{1 + \frac{\nc \snr (\nc-2)}{ \nc \snr +
\nc - 1}} - 1 \right] }.  \label{etaopt}
\end{eqnarray}
Let $\HTtrs = \frac{\etas \nc \snr}{1+\etas \nc \snr} \HT$ where 
$\HT \sim \lambda \log \left( \frac{1}{\snr} \right)$, $\kappa_1^{\star} = 
\kappa_1 |_{\etas, \hspp \HTtrs}$ and $\kappa_2^{\star} = \kappa_2 |_{\etas}$. 
If we define, 
\begin{eqnarray}
A_{1} & = &  \tsty{ \frac{ (1-\etas) ( 1 + \etas \nc \snr ) \hspp \HTtrs
\snr } {
(1 - \etas) \snr + \kappa_1^{\star} \kappa_2^{\star}} },  \\
A_{2} & = & \tsty{ \frac{ (1 -\etas)(1 + \etas \nc \snr) \hspp \HTtrs
\snr + (1 - \etas) \snr + \kappa_1^{\star} \kappa_2^{\star} } {\etas(1 - \etas)
\nc \snr^2} }, 
\end{eqnarray}
it is cumbersome, but straightforward to show that 
\begin{equation}
\lim \limits_{\snr \rightarrow 0} A_{1} = 0 \hsp \hsp {\rm and} \hsp \hsp 
\lim \limits_{\snr \rightarrow 0} \frac{1}{A_{2}} = 0
\end{equation}
for any $\mu > 0$. From (\ref{C_tr_simplify}), we then have 
\begin{eqnarray}
\max \limits_{\HTtr, \eta} 
\widehat{C}_{\train,1,\LT}(\HTtr, \eta, \nc, \snr) & \geq & 
\widehat{C}_{\train, 1, \LT}( \HTtrs, \etas, \nc , \snr) \\ 
& = & \kappa_1 \cdot \left[ \log \left( 1 +
A_{1} \right) + \nu_{A_{2}} \right]  \\
& \stackrel{(a)}{\geq} & \kappa_1 \cdot \left[ \log \left( 1 +
A_{1} \right) + \frac{1}{2} \log \left( 1+\frac{2}{A_{2}} \right)
\right] \label{xmn}
\\
& \stackrel{(b)}{\approx} & \kappa_1 \cdot \left[ A_{1} +
\frac{1}{A_{2}} \right] \label{A_approx}
\end{eqnarray}
where (a) follows from (\ref{nu_bounds}) 
and (b) is the low-$\snr$ approximation to~(\ref{xmn}). 
Substituting for $\HTtrs$ 
and simplifying
we can reduce the lower bound in
(\ref{A_approx}) to
\begin{equation}
\widehat{C}_{\train,1,\LT}(\snr)  \geq (1-\etas)
\left(\frac{\nc}{\nc-1} \right) \left(\frac{\etas \nc \snr}{1+
\etas \nc \snr} \right) \left[1 + \HT \right] \snr. 
\label{final_lb}
\end{equation}
Substituting for $\etas$ from (\ref{etaopt}) and $\nc =
\frac{1}{\snr^{\mu}}$, it can be checked that when $\mu > 1$ the
leading term is $\left[1 + \HT \right] \snr$ which equals the
first-order term of the coherent capacity as described by 
Corollary~\ref{cor1}. On the other hand when $\mu < 1$, the
leading term takes the form $\ord \left(\snr^{\frac{3-\mu}{2}}
\right)$ and hence, $\mu > 1$ is necessary.
\end{proof}
Having established the result with an average power constraint, let 
us consider the instantaneous power constraint case. 

\subsection{Achievable Rates under 
Instantaneous Power Constraint}
\label{sec4c} We impose a constraint as in (\ref{ST_const}) for the 
communication phase of the training scheme. With the same power 
allocation scheme as in (\ref{Q_inst})
(Sec.~\ref{sec3b}), we obtain
\begin{eqnarray}
\widehat{C}_{\train,1,\ST} (\snr) 
& = & \left( 1 - \frac{1}{\nc} \right)
\frac{1}{D} \hspp \sum_{i=1}^D \bEe \left[ \log \left( 1 +
\frac{ | \widehat{h}_i |^2 q_i (1 + E_{tr}) } {1 + q_i + E_{tr} }
{\hspace{0.1in}} \times \right. \right. \nonumber \\
& {\hspace{-0.7in}}& \left. \left.  \chi \left( \sum_{j = 1}^i \chi
( |\widehat{h}_j|^2 \geq \HTtr ) \leq \frac{A D
e^{-\frac{\HTtr(1+\eta \nc \snr)}{\eta \nc \snr}}}{(1-\eta)}
\right)
 \right) \right] \\
& 
= & \widehat{C}_{\train,1,\LT}(\snr) \cdot
\frac{\sum_{i=1}^{D} p^{\train}_i}{D} \label{cap_tr_final}
\end{eqnarray}
where $E_{tr} = \eta \nc \hspp \snr$ and 
$p^{\train}_i = \prob \left( \sum \limits_{j = 1}^i \chi (
|\widehat{h}_j|^2 \geq \HTtr ) \leq \frac{A D
e^{-\frac{\HTtr(1+\eta \nc \snr)}{\eta \nc \snr}}}{(1-\eta)}
\right)$. Understanding when 
$\frac{\sum_{i=1}^{D} p^{\train}_i}{D} \rightarrow 1$ is similar to the case 
studied in Sec.~\ref{sec3b}. Taking
recourse to the analysis of Prop.~\ref{prop_peak} by
using a threshold of the form $\HTtrs = \frac{\etas \nc
\snr}{1+\etas \nc \snr} \HT$ where $\etas$ is as in~(\ref{etaopt}) and 
$\HT \sim \lambda \log \left( \frac{1}{\snr} \right)$, it can be shown that the
$\frac{\sum_{i=1}^{D} p^{\train}_i}{D}$ is lower bounded by
the same expression as in (\ref{L1}) and (\ref{L2}) with $A$ replaced 
by $\frac{A}{1-\etas}$. After some simplifications, we can conclude that 
if $\frac{  \bEe \left[ \De \right]}{1 - \etas} 
- \HT \rightarrow \infty$, then $\widehat{C}_{\train,1,\ST} (\snr)
\rightarrow \widehat{C}_{\train,1,\LT} (\snr)$. Note that the condition in the 
perfect CSI case is more stringent than in the training setting. That is, 
if the channel is such that $\bEe \left[\De \right] - \HT \rightarrow \infty$, 
then it automatically ensures that 
$\frac{  \bEe \left[ \De \right]}{1 - \etas} 
- \HT \rightarrow \infty$.

\subsection{Discussion}
\label{sec4d} The analysis in Sec.~\ref{sec4b} and \ref{sec4c}
reveals that the following conditions are critical: 
\begin{enumerate}
\item[\textbf{C1})] The channel coherence dimension, $\nc$, scales with $\snr$
according to $\nc \sim \frac{1}{\snr^{\mu}}$, $\mu > 1$, and
\item[\textbf{C2})] The independent degrees of freedom (DoF), $D$, in the channel
scales with $\snr$ such that $ \frac{ \bEe \left[ \De \right]}{1 - \etas} 
- \HT = \frac{D\hspp e^{\HT }  } {1 - \etas} - \HT 
\rightarrow \infty$ as $\snr \rightarrow 0$.
\end{enumerate}
With only an average power constraint, {\bf C1} is necessary and
sufficient so that $\widehat{C}_{\train,1,\LT} (\snr) \rightarrow
\widehat{C}_{\coh,1,\LT} (\snr)$. In particular, with $\lambda
\rightarrow 1$, we approach the perfect CSI benchmark. When there is an
instantaneous power constraint, we need to satisfy \emph{both} {\bf C1} 
and {\bf C2} so that the benchmark can be attained. 

We now study the implications of these conditions. 
Note that {\bf C1} predicates a
certain minimum channel coherence level to ensure the fidelity of
the training performance. That is, the larger the value of $\mu$ and hence, 
$\nc$, the more easier it is to meet the benchmark. 
On the other hand, {\bf C2} describes the
required growth rate in the DoF, $D$, so that $\bEe \left[ \De \right] - \HT  
\rightarrow \infty$ and the instantaneous power
constraint is satisfied without any rate loss. That is, the larger the 
value of $D$, the more easier it is to meet the benchmark. It is clear
that the two conditions are somewhat conflicting in nature since
for a richer channel, it is easier to increase $D$ but more difficult 
to increase $\nc$, while for a sparser channel, it is the reverse. 
Therefore a natural question is if they can be satisfied
simultaneously.

To understand this, we first study the achievability of {\bf C1}. What are
the conditions on the channel parameters ($T_{m}$, $W_{d}$,
$\delta_1$ and $\delta_2$) and how do they interact with the
signal space parameters ($T$, $W$ and $P$) so that $\mu > 1$ is
feasible? As we discuss next, by leveraging delay and Doppler
sparsities and using peaky signaling (when necessary), $\mu > 1$ is
achievable.

\noindent \textbf{B1)} {\textbf{Rich multipath}}: When the channel
is rich in both delay and Doppler, $\nc = \frac{1}{T_{m}W_{d}}$ is
fixed and does not scale with $\snr$. Thus we can never maintain
the scaling relationship in $\nc$ as in Theorem \ref{thm2} and {\bf C1} 
can never be satisfied. Therefore, we cannot attain the benchmark
even under an average power constraint.

\noindent \textbf{B2)} {\textbf{Doppler sparsity only}}: In this
case $\Wcoh = \frac{1}{T_{m}}$ is fixed and the scaling in $\nc$
is only through $\Tcoh \sim \ftwo(T)$ (see (\ref{Ncoh})).
Therefore, by scaling $T$ with $W$ according to $ T \sim
\ftwo^{-1} \left( W^{\mu} \right)$ and choosing $\mu > 1$, we have
$\nc \sim \Tcoh \sim \ftwo \left( \ftwo^{-1} (W^{\mu}) \right)
\sim \frac{1}{\snr^{\mu}}$. For the power-law scaling in
(\ref{nc_powlaw}), we obtain
\begin{equation}
T \sim W^{\frac{\mu}{1-\delta_1}}. \label{TW_pow_dopp}
\end{equation} 
Note that as $\delta_1$ increases and the channel gets more richer, 
$T$ increases monotonically in~(\ref{TW_pow_dopp}). 

\noindent \textbf{B3)} {\textbf{Delay sparsity only}}: In this
case, $\Tcoh = \frac{1}{W_{d}}$ and $\nc = \Wcoh \Tcoh$ scales
with $\snr$ only through $\Wcoh \sim \fone \left( \frac{1}{\snr}
\right)$. Therefore, for any sub-linear function $\fone(\cdot)$, we cannot satisfy
$\mu > 1$. A possible solution to overcome this difficulty 
is to use peaky signaling where
training and communication are performed only on a subset of the
$D$ coherence subspaces. Modeling peakiness as in 
\cite{zheng,capacity_sparse_jstsp} and defining $\zeta =
\snr^{\gamma}, \hsp \gamma > 0$ as the fraction of $D$ over which
signaling is performed, it can be shown that~\cite[Lemma 3]{capacity_sparse_jstsp} 
the condition
for asymptotic coherence gets relaxed to $\nc =
\frac{1}{\snr^{\mu_{\peaky}}}$ from the original $\nc =
\frac{1}{\snr^{\mu}}$ where $\mu_{\peaky} = \mu + \gamma$. We require 
$\mu_{\peaky} > 1$ which is the same as $\mu > 1-\gamma$. For
the power-law scaling in (\ref{nc_powlaw}), we have $\nc \sim
\fone(W) \sim W^{1-\delta_2} \sim \frac{1}{\snr^{1-\delta_2}}$. Thus, 
if the peakiness coefficient $\gamma$ satisfies 
$\gamma > \delta_2$, we can satisfy the desired condition.

\noindent \textbf{B4)} {\textbf{Delay and Doppler sparsity}}:
Using (\ref{Ncoh}), we have $\Wcoh \sim \fone(W)$ and $\Tcoh \sim
\ftwo(T)$. Therefore, if we scale $T$ with $W$ according to
\begin{equation}
T  \sim  \fthr(W) \hsp \hsp \hsp \text{with} \hsp \hsp \hsp
\fthr(x) = \ftwo^{-1} \left( \frac{x^{\mu}}{\fone(x)} \right), 
\label{fthree}
\end{equation}
we have $\nc = \Wcoh \Tcoh \sim \fone(W) \ftwo(\fthr(W)) =
\fone(W) \ftwo \left( \ftwo^{-1} \left(\frac{W^{\mu}}{\fone(W)}
\right) \right) \sim \frac{1}{\snr^{\mu}}$. Thus with $\mu
> 1$ in (\ref{fthree}), we attain the desired scaling of $\nc$
with $\snr$. For the power-law scaling in (\ref{nc_powlaw}), the
desired scaling in $\nc$ can be obtained by choosing $T$, $W$ and
$P$ according to the following canonical relationship that is
obtained using (\ref{nc_powlaw}) in (\ref{fthree})
\begin{eqnarray}
T = \frac{\left( T_{m}^{\delta_2} W_{d}^{\delta_1}
\right)^{\frac{1}{1-\delta_1}} W^{\frac{\mu - 1 +
\delta_2}{1-\delta_1}}}{P^{\frac{\mu}{1-\delta_1}}}.
\label{TWP_locus}
\end{eqnarray}
From the above discussion, it is clear that channel sparsity is
necessary and in addition we also require a specific scaling
relationship between $T$ and $W$ as defined in
(\ref{TWP_locus}). But this is necessary for achieving the
benchmark capacity with an average power constraint (satisfying 
{\bf C1}). We now study how this scaling law impacts the scaling of $D$ with 
$\snr$, as in the instantaneous power case. This
is critical in determining the achievability of {\bf C2}, which we
discuss next. We recall that by definition
\begin{equation}
D = \frac{TW}{\nc} = TW \: \snr^{\mu}. \label{Dsnr1}
\end{equation}
Using (\ref{TWP_locus}) in (\ref{Dsnr1}) and simplifying,
we obtain the induced scaling behavior on $D$ with $\snr$ as
\begin{equation}
D \sim \snr^{\frac{\delta_1 \left(1-\mu \right) -
\delta_2}{1-\delta_1}}. \label{Dsnrfinal}
\end{equation}
Therefore, we have $\bEe \left[ \De \right] - \HT 
= \snr^{\lambda + \frac{\delta_1 \left(1-\mu \right) -
\delta_2}{1-\delta_1}} + \lambda \log(\snr)$ and consequently
\begin{equation}
\bEe \left[ \De \right] - \HT 
\rightarrow \begin{cases} 
\infty & {\rm if} \hsp 0 < \lambda < \frac{\delta_2 +
\left(\mu-1 \right) \delta_1}{1-\delta_1}
\\
-\infty & {\rm if} \hsp 
1 > \lambda \geq \frac{\delta_2 + \left(\mu-1 \right) \delta_1}{1-\delta_1} . 
 \end{cases}
\label{Deff_cases}
\end{equation}
It is easily seen that
\begin{equation}
\frac{\delta_2 + \left(\mu-1 \right) \delta_1}{1-\delta_1} > 1
\Longleftrightarrow \mu > \frac{1-\delta_2}{\delta_1} \end{equation}
which yields $\bEe \left[ \De \right] - \HT \rightarrow \infty$ for all
$\lambda \in (0,1)$, and {\bf C2} is satisfied as desired. The special
cases of delay sparsity only and Doppler sparsity only (as in {\bf B2} 
and {\bf B3}) are simple extensions and follow naturally.

To summarize, 
\begin{eqnarray}
\mu > 1 & \Longrightarrow & \textbf{C1 is achievable} \\
\mu > \frac{1-\delta_2}{\delta_1} & \Longrightarrow & \textbf{C2 is
achievable}. 
\end{eqnarray}
Therefore, 
\begin{equation}
\mu > \max \left( 1, \frac{1-\delta_2}{\delta_1} \right)
\Longrightarrow \textbf{C1 and C2 are achievable}. 
\end{equation}

\begin{figure}[htb!]
\begin{center}
\centerline{\includegraphics[width=4.6in]{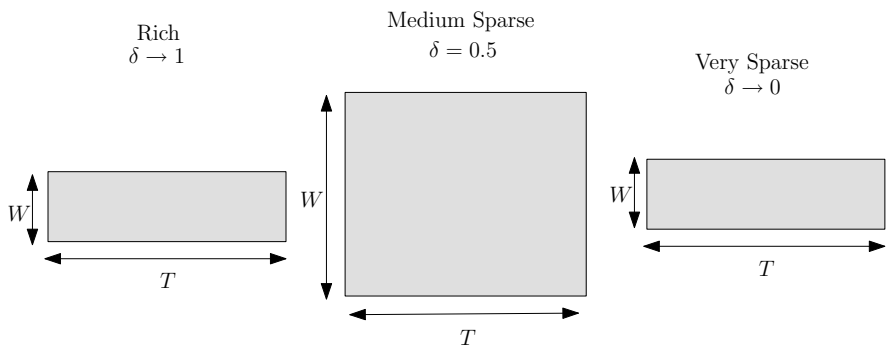}}
\end{center}
\caption{ \sl Optimal packet configurations in the non-coherent
scenario with limited feedback. Three cases illustrated here are
rich multipath $(\delta \rightarrow 1)$, medium sparsity $(\delta
= 0.5)$ and very high sparsity $(\delta \rightarrow 0)$.}
\label{fig_fb_cap_opt_pkt_tr}
\end{figure}

We now elucidate the optimal packet configurations for different levels of
channel sparsity. Analogous to the discussion in
Sec.~\ref{sec_fb_cap_opt_pkt}, we focus on the power-law
scaling and illustrate rules of thumb for choosing $T$ and $W$ 
for a given $N=TW$. Assuming symmetrical sparsity $(\delta_1 =
\delta_2 = \delta)$, we note the following two cases:
\begin{eqnarray}
\textbf{Case 1:} \: \: \frac{1-\delta}{\delta} > 1 \Longleftrightarrow
\delta < 0.5, \hsp \hsp \hsp T \sim W^{\rho}, \hsp \rho >
\frac{1-\delta}{\delta} \\
\textbf{Case 2:} \: \: \frac{1-\delta}{\delta} < 1 \Longleftrightarrow
\delta > 0.5, \hsp \hsp \hsp T \sim W^{\rho}, \hsp \rho >
\frac{\delta}{1-\delta}. 
\end{eqnarray}
The corresponding packet configurations are shown in
Fig.~\ref{fig_fb_cap_opt_pkt_tr} for $\delta \rightarrow 0$,
$\delta = 0.5$ and $\delta \rightarrow 1$. It is observed that the
slowest scaling in $T$ with $W$ is obtained for $\delta
= 0.5$ when the DoF follow a \emph{square-root} scaling law with
signal space dimension. On either extreme of this square-root
law, the required scaling in $T$ with $W$ only gets worse. This
conclusion is expected and is consistent with the contradictory
requirements presented by {\bf C1} and {\bf C2}. When $\delta < 0.5$, the
channel conditions are more favorable towards scaling $\nc$ as a
function of $\snr$ (specified by {\bf C1}). However, the required
scaling of $D$ with $\snr$ (specified by {\bf C2}) is non-trivial and
ultimately dominates the required scaling of $T$ with $W$. On the
other hand, when $\delta > 0.5$, the relatively less sparse
channel conditions are favorably disposed towards the scaling of
$D$ as a function of $\snr$, but this is at the cost of scaling in
$\nc$. For the case of asymmetrically sparse channels, it can
be shown that this desirable condition (slowest scaling of $T$
with $W$) generalizes to $\delta_1 + \delta_2 = 1$.
{\vspace{-5mm}}

\section{Concluding Remarks}
\label{sec5} In this paper, we studied the achievable rates of
sparse multipath channels with limited feedback. The focus of our
analysis is in the wideband$/$low-$\snr$ regime. Our investigation
includes constraining both the average and the instantaneous
transmit powers. We first analyzed the case when the receiver has
perfect CSI and when one bit (per channel coefficient) 
of this CSI is known perfectly at the transmitter. 
We established conditions under which the
rates achievable with this scheme approach the capacity with
perfect receiver and transmitter CSI. For sparse channels, these 
conditions translate to certain optimal 
packet configurations for signaling. 
When the receiver has no CSI \emph{a priori}, we
studied the performance of a training scheme.
It is shown that with only an average power constraint, channel
sparsity is necessary to attain the coherent
performance. With an instataneous power constraint, we established 
conditions on optimal packet
configurations in order to approach the benchmark capacity gain asymptotically 
as $\snr \rightarrow 0$.

\begin{table}[htb!]
\caption{Conditions necessary to achieve the perfect CSI benchmark of 
$\log \left( \frac{1}{\snr} \right ) \snr$.}
\begin{center}
\begin{tabular}{|c|c|c|c|c|}
\hline \hline 
CSI & CSI & Power  & Necessary & Signaling \\ 
Rx.\ & Tx.\ & Const.\ & Conditions & Parameters 
\\ 
\hline \hline 
Perf.\ & Perf.\ & - &  $h_w \sim \log \left(  \frac{1}{\snr} \right)$ & 
Waterfilling; see~\cite{verdu,borade} \\ 
Perf.\ & $1$ bit & Avg.\ & 
$\HT = \lambda \log \left( \frac{1}{\snr} \right),\hsp \lambda \rightarrow 1$ 
& No constraints on richness or $T$, $W$;
\\ 
& & & & see~\cite{verdu,borade,manish}
\\
Perf.\ & $1$ bit & Inst.\ & 
$\HT =  \lambda \log \left( \frac{1}{\snr} \right )$ 
& Rich channel: no constraint on $T$ or $W$, \\ 
& & & for $\lambda < 1$, and & 
Sparse ($T$ fixed): $\lambda < \delta_2$ limits rates, \\ 
& & & $\bEe \left[ \De \right] - \HT \rightarrow \infty $  & 
Sparse (general): $T \sim W^{\rho}$, \hsppp 
$\rho \geq \frac{1 - \delta_2}{\delta_1}$ \\ 
Train.\ & $1$ bit 
& Avg.\ & $\nc \sim \frac{1}{\snr^{\mu}}, \hsp \mu > 1$ 
& Rich channel: Impossible, \\ 
& & & & Sparsity (Doppler): Non-peaky  \\ 
& && & scheme with $T \sim W^{ \frac{\mu} {1 - \delta_1} }$, \\ 
& & & & 
Sparsity (delay): Peaky scheme with \\ 
& && & peakiness coefficient $\gamma > \delta_2$, 
\\ 
&& & & Sparsity (both): Non-peaky scheme; \\ 
& & & & see~(\ref{fthree}) and~(\ref{TWP_locus}) \\ 
Train.\ & $1$ bit 
& Inst.\ & $\nc \sim \frac{1}{\snr^{\mu}}, \hsp \mu > 1$ 
& Rich channel: Impossible, \\
& & & and $\frac{\bEe \left[ \De \right] }{1 - \etas} 
- \HT \rightarrow \infty$  &  
Sparse (both): $\mu > \frac{1 - \delta_2}{\delta_1}$ for no rate 
\\ 
& & & & 
loss, else 
$\lambda < \frac{ \delta_2 + (\mu - 1) \delta_1}{ 1 - \delta_1}$ 
\\ &&&&
\\ \hline 
\end{tabular}
\label{tabla1}
\end{center}
\end{table}

We contrast the results of this work with recent 
observations in \cite{borade,manish}. The focus in \cite{borade,manish} 
is on training schemes and on scenarios where
$\Tcoh$ increases as $\snr$ decreases, although there is no
mention of how such a scaling law can be realized in practice. In
particular, the authors show that capacity scales as $\log \left(
\Tcoh \right) \snr$ if $\log(\Tcoh) \preceq \log
\left(\frac{1}{\snr} \right)$ and equals the coherent capacity,
$\log \left(\frac{1}{\snr} \right) \snr$, when $\log(\Tcoh) \succeq
\log \left( \frac{1}{\snr} \right)$. On the other hand, we have
shown that when the channel is sparse, channel coherence scales
naturally with $T$ and $W$ and the benchmark gain, $\log \left(
\frac{1}{\snr} \right)$, can always be achieved by appropriately
choosing $T$ and $W$. Furthermore, while \cite{borade,manish}
considered only an average power constraint, we have established
achievability under both average and instantaneous power
constraints. Also, peaky training schemes are necessary in the
framework of \cite{borade} to achieve perfect training
performance. Such schemes would violate any finite instantaneous
power constraint. Our findings here reveal that channel sparsity
is a degree of freedom that can be exploited to obtain
near-coherent performance with non-peaky training schemes. 
Table~\ref{tabla1} provides a short summary of our contributions and 
places them in the context of~\cite{verdu,borade,manish}.

Finally, we note that the results obtained here closely parallel
our earlier work~\cite{capacity_sparse_jstsp} where we studied the 
achievable rates with training and no feedback. We showed that 
when
$\nc = \frac{1}{\snr^{\mu}}$ with $\mu > 1$, the channel is
\emph{asymptotically coherent}; channel estimation performance is
near-perfect at a vanishing energy cost. Analogous to~\cite{capacity_sparse_jstsp}, 
we have shown here that under the
assumption of an error-free $D$-bit feedback link, the rate achievable with 
the training scheme converges to the perfect CSI 
benchmark. Furthermore,
the cost of feedback, measured in terms of the number of feedback
bits per dimension $(D/N)$ converges asymptotically to zero in a sparse 
channel.


\appendix
\subsection{Tightness of $\widehat{C}_{\coh,1,\LT}(\snr)$ to $C_{\coh,1,\LT}(\snr)$ 
as $\snr \rightarrow 0$}
\label{app_closeness} Let $\chi_i$ denote the random variable
$\chi( | h_i|^2 \geq \HT )$. Defining $\gamma
\triangleq \frac{ \left| C_{\coh,1,\LT}(\snr) -
\widehat{C}_{\coh,1,\LT}(\snr) \right|  } { C_{\coh,1,\LT}(\snr) }$, we have
\begin{eqnarray}
\gamma & = & \frac{1}{D} \left| \sum_{i=1}^D \bEe \left[ \log
\left( 1 + \frac{ \frac{TP | h_i|^2 \chi_i (D e^{-\HT} - \sum_{i}
\chi_i )  }{ \sum_i \chi_i N_c D e^{-\HT}  }  } { 1 + \frac{ TP
| h_i|^2 \chi_i }
{N_c D e^{-\HT}}  } \right) \right] \right|  \\
& \leq & \frac{1}{D}  \sum_{i=1}^D \bEe \left| \log \left( 1 +
\frac{ \frac{TP | h_i|^2 \chi_i (D e^{- \HT} - \sum_{i} \chi_i )
}{ \sum_i \chi_i N_c D e^{-\HT}  }  } { 1 + \frac{ TP | h_i|^2
\chi_i }
{N_c D e^{-\HT}}  } \right) \right|  \\
& \stackrel{(a)}{\leq} & \frac{1}{D} \sum_{i=1}^D \bEe \left|
\frac{ \frac{TP | h_i|^2 \chi_i (D e^{- \HT} - \sum_{i} \chi_i )
}{ \sum_i \chi_i N_c D e^{-\HT}  }  } { 1 + \frac{ TP | h_i|^2
\chi_i }
{N_c D e^{-\HT}}  }  \right|  \\
& = & \frac{TP} {N_c D^2 e^{-\HT}} \sum_{i=1}^D \bEe \left[ \frac{
| h_i|^2 \chi_i \left|De^{-\HT} -\sum_i\chi_i \right| } {\sum_i
\chi_i \left( 1+ \frac{TP |h_i|^2 \chi_i }{ N_c D e^{-\HT} }
\right) }
\right]  \\
& \stackrel{(b)}{ =} & \frac{TP} {N_c D e^{-\HT}} \bEe \left[
\frac{ |h_1|^2 \chi_1 \left|De^{-\HT} -\sum_i\chi_i \right| }
{\sum_i \chi_i \left( 1+ \frac{TP |h_1|^2 \chi_1 }{ N_c D
e^{-\HT} }  \right) } \right] \triangleq \gamma_0
\end{eqnarray}
where (a) follows from the log-inequality and (b) from the fact
that $\{ h_i \}$ are i.i.d. Conditioning on $\chi_1$, we now have
\begin{eqnarray}
\gamma_0 & = & \frac{TP} {N_c D e^{-\HT}} \bEe[\chi_1] \hspp
\bEe_{h_1, \{ \chi_j, j>1 \}} \left[ \frac{ |h_1|^2
\left|De^{-\HT} - (1 + \sum_{j>1}\chi_j) \right| } { (1 + \sum_{j
> 1} \chi_j)
\left( 1+ \frac{TP |h_1|^2  }{ N_c D e^{-\HT} }  \right) } \right] \\
& = & \snr \cdot \bEe_{h_1, \{ \chi_j, j>1 \}} \left[ \frac{
|h_1|^2  \left|De^{-\HT} - (1 + \sum_{j>1}\chi_j) \right| } { (1
+ \sum_{j > 1} \chi_j) \left( 1+ \frac{TP |h_1|^2  }{ N_c D
e^{-\HT} }  \right) } \right]
\\
& \stackrel{(a)}{=} & \snr \cdot \bEe_{ h_1 } \left[ \frac{
|h_1|^2 } { 1+ \frac{TP |h_1|^2  }{ N_c D e^{-\HT} } } \right]
\cdot \bEe_{ \{ \chi_j, j > 1 \} } \left[ \frac{\left|De^{-\HT} -
(1 + \sum_{j>1}\chi_j) \right|}
{ (1 + \sum_{j > 1} \chi_j)   } \right]  \\
& \leq & \snr \cdot \bEe[ |h_1|^2 ] \cdot \bEe_{ \{ \chi_j, j >
1 \} } \left[ \frac{\left|De^{-\HT} - (1 + \sum_{j>1}\chi_j)
\right|} { (1 + \sum_{j > 1} \chi_j)   } \right] \triangleq
\gamma_1
\end{eqnarray}
where (a) follows from the fact that $h_1$ and $\{ \chi_j, j > 1
\}$ are independent.

To show the closeness of $\widehat{C}_{\coh,1,\LT}(\snr)$ to
$C_{\coh,1,\LT}(\snr)$, we now produce an upper bound for $\gamma_1$
that tends to $0$ as $\snr \rightarrow 0$. Our goal is to show
that given any choice of $D$, $\frac{\gamma_1}{\snr}$ is bounded.
Consider
\begin{eqnarray}
\bEe_{ \{ \chi_j, j > 1 \} } \left[ \frac{\left|De^{-\HT} - (1 +
\sum_{j>1}\chi_j) \right|} { (1 + \sum_{j > 1} \chi_j)   } \right]
& = & \bEe_{ \{ \chi_j, j > 1 \}  } \left[ \left| \frac{ De^{-\HT}
}{ (1 + \sum_{j > 1} \chi_j) }
- 1 \right| \right] \nonumber \\
& \stackrel{(a)}{\leq} & \underbrace{ \sqrt{ \bEe_{\chi_j} \left[
\left( \frac{ De^{-\HT} }{ (1 + \sum_{j > 1} \chi_j) } \right)^2 +
1 - 2 \frac{ De^{-\HT} }{ (1 + \sum_{j > 1} \chi_j) } \right] } }_
{\triangleq \gamma_2 }  \nonumber
\end{eqnarray}
where (a) is a consequence of Cauchy-Schwarz inequality. Let $\ee$
denote $e^{-\HT}$. We then have
\begin{eqnarray}
\gamma_2 \stackrel{(b)}{\leq} \sqrt{ 1 + D^2 \ee^2 \cdot
\bEe_{\chi_j} \left[ \frac{1}{ (1 + \sum_{j > 1} \chi_j)^2}
\right] - \frac{2 D \ee }{1 + (D-1)\ee} }
\end{eqnarray}
where in (b) we have used the fact that $\bEe\left[ \frac{1}{\bX}
\right] \geq \frac{1}{\bEe[\bX]}$ for a positive random variable
$\bX$. We now estimate $\alpha \triangleq \bEe_{\chi_j} \left[
\frac{1}{ (1 + \sum_{j > 1} \chi_j)^2}  \right]$. It is easy to
check that
\begin{eqnarray}
\alpha = \sum_{i=0}^{D-1} { D-1 \choose i} \frac{\ee^i
(1- \ee)^{D-1-i}} { (i+1)^2 } .
\end{eqnarray}
Noting that
\begin{eqnarray}
\label{one} (1+y)^{D-1} = \sum_{i=0}^{D-1} { D-1 \choose i} \hsp y^i
\end{eqnarray}
and integrating twice both sides of (\ref{one}) with respect to
$y$, we have
\begin{eqnarray}
\label{two} \frac{(1+y)^{D+1}} { D(D+1) } = \sum_{i=0}^{D-1} { D-1
\choose i} \hsp \frac{ y^{i+2} }{ (i+1)(i+2) }.
\end{eqnarray}
Using $y = \frac{\ee}{1 - \ee}$ in (\ref{two}), we have
\begin{eqnarray}
\frac{ 1}{ D(D+1) \ee^2 } = \sum_{i=0}^{D-1} { D-1 \choose i}
\frac{\ee^i (1 - \ee)^{D-1-i} }{(i+1)(i+2)}.
\end{eqnarray}
Observe that $\frac{1}{(i+1)^2} \leq \frac{2}{ (i+1)(i+2) }$ for
all $i \geq 0$ and an upper bound for $\gamma_2$ is
\begin{eqnarray}
\label{ubd_f} \gamma_2 \leq \sqrt{ 1 +\frac{2 D^2 \ee^2} {D (D+1) \ee^2}
- \frac{2 D \ee }{1 + (D - 1) \ee} } = \sqrt{ \frac{D^2 \ee -4 D \ee +3 D - \ee + 1}
{(D + 1)(D \ee - \ee + 1) } }
\end{eqnarray}
which is bounded for any choice of $D$. (In fact, the upper bound
converges to $1$ as $D \rightarrow \infty$). Note that the bound
in (\ref{ubd_f}) is loose and one might expect that
$\frac{\gamma_1}{\snr} \rightarrow 0$ as $D \rightarrow \infty$ as
a consequence of the law of large numbers. However, for our
purpose, the proposed loose upper bound in (\ref{ubd_f}) is
sufficient.

\subsection{Proof of Proposition~\ref{prop_peak}}
\label{app_ltpc}
To compute $p_i \triangleq \prob \left(
\sum_{j = 1}^i \chi ( |h_j|^2 \geq \HT ) \leq A D e^{-\HT} \right)$, we need the
following result~\cite[Theorem 2.8,~p.\ 57]{petrov} on
the tail probability of a sum of independent random variables.
\begin{lem}
\label{lem_petrov} Let $\bX_i, i = 1 , \cdots ,n$ be independent
random variables with $\bEe[\bX_i] = 0$ and $\bEe[ \bX_i^2] =
\sigma_i^2$. Define $B_n = \sum_{i=1}^n \sigma_i^2$. If there
exists a positive constant $H$ such that
\begin{eqnarray}
\bEe[ \bX_i^m ] \leq \frac{1}{2} m! \sigma_i^2 H^{m-2}
\end{eqnarray}
for all $i$ and $x \geq \frac{B_n}{H}$, then we have $\prob \big(
\sum_{i=1}^n \bX_i > x \big) \leq \exp \big( -\frac{x}{4H} \big)$. If $x \leq
\frac{B_n}{H}$, then we have $\prob \big( \sum_{i=1}^n \bX_i > x \big) \leq
\exp \big( -\frac{x^2}{4B_n} \big)$.
\endproof
\end{lem}
\noindent
To apply Lemma~\ref{lem_petrov}, we set $n = i$ and $\bX_j = \chi( |h_j|^2
\geq \HT) - \bEe \left[ \chi( |h_j|^2 \geq \HT) \right] = \chi( |h_j|^2 \geq \HT) -
e^{-\HT} = \chi_j - \ee$ for $j = 1 , \cdots, i$.
Then, a simple computation of the higher moments of $\bX_j$ implies that
$\bEe[\bX_j^2 ] = \sigma_j^2 = \ee (1 - \ee)$, $B_i = i \ee (1 - \ee)$,
$\bEe[ \bX_j^m ] = \ee (1 - \ee) \cdot (  (1 - \ee)^{m-1} + (-1)^m \ee^{m-1} )$.
It can be checked that $H = (1 - \ee)$ is sufficient to satisfy the conditions
of Lemma~\ref{lem_petrov}. With this setting, we have
\begin{eqnarray}
\prob \left(  \sum_{j=1}^i \chi( |h_j|^2 \geq \HT ) - i\ee >
(AD-i)\ee \right) 
\leq \left\{
\begin{array}{cc}
\exp \left( - \frac{ (AD- i) \ee}{4(1-\ee)}  \right)
& {\rm if} \hsp i \leq \lfloor \frac{AD}{2} \rfloor, \\
\exp \left( - \frac{ (AD - i)^2 \ee  }{ 4 i (1 - \ee) } \right)
& {\rm if} \hsp i \geq \lfloor \frac{AD}{2} \rfloor +1.
\end{array}
\right. \label{prob_eqn}
\end{eqnarray}

If $1 < A < 2$, with $\kappa = \frac{\ee}{4(1 - \ee)}$ using~(\ref{prob_eqn}), the
following lower bound, $L$, holds for $\frac{ \sum_{i=1}^D p_i}{D}$:
\begin{eqnarray}
L & = &  1 - \left[ e^{-AD \kappa} \sum_{ i \leq \lfloor \frac{AD}{2}
\rfloor } e^{i \kappa} + \sum_{ i \geq \lfloor \frac{AD}{2}
\rfloor + 1 } e^{ - \frac{ (AD - i)^2 \kappa }{ i  } } \right]
\\ 
& \stackrel{(a)}{=} & 
1 - \left[ \frac{ e^{- \kappa (AD - 1)} \cdot
(e^{\kappa \lfloor \frac{AD}{2} \rfloor}  - 1)  } { e^{\kappa} -
1}  
+ \left(D -
\left\lfloor \frac{AD}{2} \right\rfloor \right) e^{- (A-1)^2 D
\kappa} \right]  \\ 
& \geq & 1 - \left[ \frac{1} { e^{\kappa } - 1 } \cdot e^{-
\kappa\left(\frac{ AD}{2} -1\right) }  
+ \left( 1+ D (1 - A/2 ) \right) e^{-
(A-1)^2 D \kappa} \right] \label{1A2}
\end{eqnarray}
where (a) follows by first using $\frac{(AD - i)^2}{i} \geq
(A-1)^2 D$ for all $1 \leq i \leq D$ and then upon further
simplification using the sum of a geometric series.

If $A \geq 2$, we have the following lower bound to $\frac{
\sum_{i=1}^D p_i}{D}$:
\begin{equation}
L  =   1 -  \exp(-AD \kappa) \sum_{1 \leq i \leq D } e^{i \kappa}
\approx 1 - e^{ -\kappa \left( D(A - 1) - 1 \right)} \cdot
\frac{1}{e^{\kappa} - 1} \label{2A2}.
\end{equation}

With $\HT = \lambda \log \left( \frac{1}{\snr}  \right)$ as
in~(\ref{HT_eqn}), the dominant term of $\ee$ is $\snr^{\lambda}$
and hence in $\kappa$ is $\frac{\snr^{\lambda}}{4}$. With this
choice of $\HT$ in (\ref{1A2}) and (\ref{2A2}) and simplifying, we
obtain the desired bounds in (\ref{L1}) and (\ref{L2}). 
It is also straightforward to check that when $D$ satisfies $D \hspp
\snr^{\lambda} + \lambda \log(\snr) \rightarrow \infty$ as $\snr \rightarrow 0$, $L
\rightarrow 1$ in both the cases.
\endproof

\subsection{Completing the Proof of Theorem~\ref{thm2}}
\label{app_pf_thm2} The choice of $\HT$ we study is $\HT = \ep
\log \left( \frac{1}{\snr} \right)$ for some $\ep > 0$. First,
with this fixed choice of $\HT$, note that maximizing
$\widehat{C}_{\train,1,\LT} \left( \eta, \nc, \snr \right)$ is
equivalent to setting its derivative (with respect to $\eta$) to
zero. Then, it is straightforward to check that the derivative is
\begin{eqnarray}
&& \underbrace{\frac{\nu_{\beta} \HT}{\eta} }_{{\sf I}} +
\underbrace{ \frac{\HT}{\eta } \log_e \left( 1  +
\frac{(1- \eta) (1 + \eta \nc \snr) \HT \snr }
{(1 - \eta) \snr + \kappa_1 \kappa_2} \right) }_{ {\sf II}} \nonumber \\
& + & \underbrace{ \frac{\left(\nu_{\beta} - \frac{1}{\beta} \right)}
{\snr \eta}
\left[ \kappa_1 \left(1 - \frac{1}{\nc}\right)
\left( \frac{\nc \eta^2 \snr + 2 \eta - 1}{(1 - \eta)^2} +
\frac{ \HT (1 + \eta \nc \snr)}{\eta \nc \snr (1- \eta)} \right)
- \snr(\HT+1) \right] }_{ {\sf III}}  \nonumber \\
&+& \underbrace{
\frac{\HT \snr^2 \nc \eta}{ (1 - \eta) \snr + \kappa_1 \kappa_2 }
\cdot
\frac{ \nc \snr^2 (1 - \eta)^2 - \kappa_1 \kappa_2( 1 + \eta \snr \nc )
\left(1 + \frac{\HT (1 - \eta)}{\nc \eta^2 \snr}  \right)}
{(1 - \eta) \snr + \kappa_1 \kappa_2 + (1 - \eta)(1 + \eta \nc \snr) \HT \snr  } }_{
{\sf IV}}.
\label{derivative}
\end{eqnarray}
For simplicity, we will denote the four terms in~(\ref{derivative}) by {\sf I},
{\sf II}, {\sf III} and {\sf IV}. We will further assume that
$\eta = \snr^x, x \geq 0$ and $\nc = \frac{1}{\snr^y}, y > 0$.
For a given choice of $\ep$, our goal is to determine the relationship between
$x$ and $y$ such that the derivative in~(\ref{derivative}) can be zero.
We consider three cases: i) $y > 1 + x$, ii) $y < 1 + x$ and iii) $y = 1 + x$.

\noindent {\bf \em Case i:}
First, note that $\eta \nc \snr = \snr^{-z}$ for some $z > 0$. The dominant terms
of $\beta$ can be seen to be $\frac{1}{\snr^{1-\ep}} + \ep \log \left(
\frac{1}{\snr}\right)$ and thus, up to first order $\beta = \frac{1}{\snr^{1-\ep}}$.
Similarly, $(1 - \eta) \snr + \kappa_1 \kappa_2$ up to first order equals
$\snr^{\ep-z}$. Note from~\cite[5.1.20, p.\ 229]{stegun} that
$\nu_{\beta} = \ord \left( \frac{1}{\beta} \right)$ if $\beta \rightarrow \infty$
and hence {\sf I} is $\ep
\log \left( \frac{1}{\snr} \right) \frac{1}{\snr^{\ep + x - 1}}$. It can also be
checked that {\sf II} is $\left( \ep \log \left( \frac{1}{\snr} \right)\right)^2
\frac{1}{\snr^{ \ep + x - 1}}$, $\nu_{\beta} - \frac{1}{\beta} =
\ord \left( \frac{1}{\beta^2}\right)$ and hence {\sf III} is $\ep \log \left(
\frac{1}{\snr} \right) \frac{1}{\snr^{\ep + x - 1}}$ as long as $y < 1 + 2x$. Under
the same assumption, $y < 1 + 2x$, {\sf IV} is $- \left(\ep \log \left(
\frac{1}{\snr}\right) \right)^2 \frac{1}{\snr^{\ep + x - 1}}$. Thus, by playing
with constants the derivative can be set to zero in this case. If $y \geq 1 + 2x$,
{\sf I} and {\sf II} remain unchanged, but {\sf III} is $\snr^{2 + x - y - \ep}$
and {\sf IV} is $-\ep \log \left( \frac{1}{\snr} \right)\snr^{2 + x - y - \ep}$.
By comparing the coefficients, we see that the only way the derivative can be zero
is if $y = 1 + 2x$.

\noindent {\bf \em Case ii:}
In this case, the first order terms show the following behavior.
With $w = 1 + x - y > 0$, {\sf I} is $\snr^{w  - x}$, {\sf II} is $\ep
\log \left( \frac{1}{\snr}\right) \log \log \left(\frac{1}{\snr} \right)
\frac{1}{\snr^x}$, {\sf III} is $- \snr^{2 w - x} \frac{1}{\ep \log
\left(\frac{1}{\snr}\right)}$, and {\sf IV} is $\snr^{2 - 2 y + x}$.
It can be seen that the derivative can never be zero and hence this case is
ruled out.

\noindent {\bf \em Case iii:}
In this case, based on a similar analysis, we see that the derivative can
again be set to zero.

Therefore, if $\ep \in (0,1)$, $x \geq 0$ and $1+ x < y \leq 1 +
2x$, we have
\begin{eqnarray}
\widehat{C}_{\train,1,\LT}(\snr) \geq \snr^{\ep} \log \left( 1 +
\frac{\ep \log \left( \frac{1}{\snr} \right) \snr^{1 - \ep} (1 -
\snr^x) } {1 - \snr^y} \right) + \snr.
\end{eqnarray}
Thus, $\widehat{C}_{\train,1,\LT}(\snr)$ is up to first order the
same as $\widehat{C}_{\coh,1,\LT}(\snr)$ and
$C_{\coh,1,\LT}(\snr)$. If $y = 1 + x$ and $\eta \nc \snr = a$ for
some choice of $a$ (positive, finite and independent of $\snr$),
we need $a > \frac{\ep}{1 - \ep}$ and we have
\begin{eqnarray}
\widehat{C}_{\train,1,\LT}(\snr) \geq \snr^{\frac{\ep(1+a)}{a} }
\log \left( 1 + \ep \snr^{1 - \frac{\ep(1 +a)}{a}} \log \left(
\frac{1}{\snr} \right )  \right) + \frac{a}{1 + a} \cdot \snr.
\end{eqnarray}
If $y < 1+x$, the training scheme is strictly sub-optimal (in the
limit of $\snr$) from an ergodic capacity point-of-view. Putting
things together, we obtain the desired condition, $\mu > 1$.
\endproof

\bibliographystyle{IEEEbib}
\bibliography{bib_jstsp2}

\end{document}